%% file: DGVZ.tex
\documentclass[12pt]{article}
\usepackage[utf8]{inputenc}
  \usepackage{graphicx}
  \usepackage{epstopdf} 
  \usepackage{epsfig}
  \usepackage{amsmath}
  \usepackage{amssymb}
  \usepackage{subcaption}
  \usepackage{color}
  \usepackage{verbatim}
\usepackage{multirow}
\usepackage[bottom]{footmisc}
\usepackage{cite,url}
\usepackage{hyperref}
\usepackage[normalem]{ulem}

\evensidemargin - .5truecm
\newcommand\amp{{\cal A}}
\newcommand{\be}{\begin{equation}}
\newcommand{\ee}{\end{equation}}
\newcommand{\nn}{\nonumber}
\newcommand{\bea}{\begin{eqnarray}}
\newcommand{\eea}{\end{eqnarray}}
\newcommand{\bfig}{\begin{figure}}
\newcommand{\efig}{\end{figure}}
\newcommand{\bc}{\begin{center}}
\newcommand{\ec}{\end{center}}

\def\as{\alpha_s}

\def\sq2{\sqrt{2}}
\newcommand{\smallz}{{\scriptscriptstyle Z}} %
\newcommand{\mz}{m_\smallz}

\newcommand{\smallh}{{\scriptscriptstyle H}}
\newcommand{\mh}{m_\smallh}
\newcommand{\mt}{m_t}

\newcommand{\Cnlo}{{\cal C}_{\scriptscriptstyle{\rm NLO}}}

\newcommand{\MSbar}{\overline{\mathrm{MS}}}

\begin{document}
\begin{titlepage}
\nopagebreak
{\flushright{
        \begin{minipage}{5cm}
         CERN-TH-2022-079
        \end{minipage}        }

}
\renewcommand{\thefootnote}{\fnsymbol{footnote}}
\vspace{1cm}
\begin{center}
  {\Large \bf \color{magenta} On the NLO QCD Corrections to Gluon-Initiated
    $ZH$ Production}
  
\bigskip\color{black}\vspace{0.6cm}
     {\large\bf  Giuseppe Degrassi$^{a,b}$\footnote{email: giuseppe.degrassi@uniroma3.it},
       Ramona Gr\"{o}ber$^{c}$\footnote{email: ramona.groeber@pd.infn.it},
       Marco Vitti$^b$\footnote{email: marco.vitti@uniroma3.it} and
       Xiaoran Zhao$^b$\footnote{email: xiaoran.zhao@uniroma3.it}}
     \\[7mm]
{\it  (a) CERN, Theoretical Physics Department, 1211 Geneva 23,
         Switzerland}\\[1mm]
 {\it (b) Dipartimento di Matematica e Fisica, Universit{\`a} di Roma Tre and \\
 INFN, sezione di Roma Tre, I-00146 Rome, Italy}\\[1mm]
{\it (c) Dipartimento di Fisica e Astronomia 'G.~Galilei', Universit\`a di Padova and INFN, sezione di Padova, I-35131 Padova, Italy}\\     
\end{center}

\bigskip
\bigskip
\bigskip
\vspace{0.cm}

\begin{abstract}
We compute the QCD corrections at next-to-leading order for the
process $gg \rightarrow ZH$, including both the virtual two-loop terms
and real-emission contributions. The two-loop box diagrams in the
virtual corrections are approximated analytically over the complete
phase space, combining the results of an expansion in the limit of 
small transverse momentum and an expansion in the regime of high energy. We
obtain both inclusive and differential results for the cross
section. We find that the NLO QCD corrections are of the same size as
the LO contribution up to $ZH$ invariant masses close to 1 TeV, but
they increase significantly when higher energies are considered, due
to a class of real-emission diagrams in which the $Z$ boson is
radiated from an open quark line. Finally, we estimate the uncertainty due to
the renormalization scheme used for the top-quark mass both on the total and
differential cross section.  

\end{abstract}
\vfill  
\end{titlepage}    

\setcounter{footnote}{0}

\input{DGVZIntro}

\input{DGVZ2}

\input{Res}

\input{Concl}

\section*{Acknowledgements}
We thank Luigi Bellafronte and Pier Paolo Giardino for useful discussions.
The work of M.V.~and X.Z.~is supported by the Italian Ministry of Research (MUR) under grant PRIN 20172LNEEZ. 

\bibliographystyle{utphys}
\bibliography{ggzh_nlo}

\end{document}

%% file: DGVZIntro.tex
\section{Introduction}
In the last ten years following the discovery of the Higgs boson
\cite{ATLAS:2012yve, CMS:2012qbp} a great experimental effort has been
undertaken at the Large Hadron Collider (LHC) in order to measure the
properties of this particle with high precision. The increased statistics expected
from the next LHC runs and from the High-Luminosity phase will allow
for more accurate comparisons with the Standard Model (SM) predictions, and
refined theoretical calculations are needed to extract as much
information as possible from the forthcoming measurements\footnote{A
comprehensive review on precision calculations for Higgs physics can
be found in Ref.\cite{Heinrich:2020ybq}}.

The process in which a Higgs is produced together with a weak vector
boson, known as $VH$ associated production ($V=W,Z$), is of great
relevance at the LHC not only as a probe of the couplings between the
Higgs and the weak bosons, but also because of its sensitivity to the
$H \rightarrow b \bar{b}$ decay. Indeed, the large QCD backgrounds
affecting searches of $H \rightarrow b \bar{b}$ can be reduced more
efficiently in $VH$ than in other production modes
\cite{ATLAS:2018kot, CMS:2018nsn, ATLAS:2020fcp}.  Recently $VH$
production has been considered also to improve the constraints on the
charm Yukawa coupling \cite{ATLAS:2022ers}.

In current $VH$ analyses the impact of theoretical uncertainties from
missing higher-order terms in perturbative calculations depends on the
choice of the final-state boson $V$: while in the $WH$ case these
uncertainties are around 1\%, and are comparable to the uncertainties
from the PDF and $\as$ determination, the corresponding
uncertainties for $ZH$ production are slightly larger, close to 3\%
\cite{LHCHXSWG:2016ypw}. In view of the progress in experimental precision it is hence essential
to improve the theoretical
control over $pp \rightarrow ZH$.

There are two partonic channels that contribute to the $pp \rightarrow
ZH$ hadronic cross section. The $q\bar{q}$-initiated (Drell-Yan-like) channel gives the leading contribution, and higher-order
corrections are well under control as they are known to
next-to-next-to-leading order (NNLO) in QCD \cite{Han:1991ia,
  Brein:2003wg, Brein:2011vx} and to NLO in the EW interactions
\cite{Ciccolini:2003jy, Denner:2011id}. The $gg$-initiated channel has
been computed at LO in Refs.~\cite{Kniehl:1990iva, Dicus:1988yh}, and
it contributes for the first time as a NNLO QCD correction to the
hadronic process. However, this sub-leading channel is enhanced by the
large gluon luminosity at the LHC, and it provides about 6\% of the
total cross section for 13 TeV collisions. Differential analyses have
pointed to significant differences in the shapes of distributions
between the $q \bar{q}$ and the $gg$ channels \cite{Ferrera:2014lca,
  Hespel:2015zea, Harlander:2018yio, Bizon:2021rww}, showing an increased  relative importance
  of the $gg$ channel in the boosted regime \cite{Englert:2013vua}.  The
$gg\rightarrow ZH$ process has been considered also as a probe of new
physics effects, with examples including anomalous
couplings\cite{Bizon:2021rww, Englert:2013vua, Englert:2016hvy,
  Yan:2021veo} and new degrees of freedom \cite{Kniehl:2011aa,
  Englert:2013vua}. Finally, the gluon-initiated channel has the largest
impact on the theoretical uncertainties for $pp \to ZH$, as only the
LO contribution is included in the Monte Carlo programs used for
experimental studies. Since an equivalent gluon-initiated contribution
is not allowed in $WH$ production because of electric charge
conservation, $gg \rightarrow ZH$ is responsible for the larger
uncertainties in $ZH$ production compared to $WH$. The reduction
of the current theoretical uncertainties requires the calculation of
the NLO QCD corrections to $gg \rightarrow ZH$, which are the main
subject of this paper.

Since the main contribution to $gg \to ZH$ at LO comes from one-loop diagrams involving loops of massive quarks, the most challenging part of the NLO
calculation is associated to two-loop multi-scale integrals in the
virtual corrections. An estimate of these terms has been obtained in
Ref.~\cite{Altenkamp:2012sx} in the $m_t \rightarrow \infty$ limit and
included in the program $\texttt{vh@nnlo}$ \cite{Brein:2012ne,
  Harlander:2018yio}, while the effects of a large but finite
top-quark mass have been considered in Refs.~\cite{Hasselhuhn:2016rqt,
  Grober:2017uho}. An approximation that retains the effects of the
top-quark mass and that is accurate over the complete phase space has been
presented in Ref.~\cite{Wang:2021rxu}, whereas the very recent
calculation of Ref.~\cite{Chen:2022rua} relies on a combination of the
Pad\'e-improved high-energy expansion presented in Ref.~\cite{Davies:2020drs} and a numerical evaluation using
sector decomposition \cite{Chen:2020gaew}. All of the previous
results show that the NLO corrections are comparable in magnitude to
the LO contribution\footnote{In fact, the NLO corrections can become
substantially larger than the LO contribution in transverse-momentum
distributions \cite{Chen:2022rua}.}, and this fact makes the
implementation of the NLO terms in a Monte Carlo code (possibly
interfaced with parton-shower generators \cite{Alioli:2019qzz, Zanoli:2021iyp}) a
priority for a reliable interpretation of the experimental
results\footnote{Additionally, in Ref.~\cite{Harlander:2014wda} it has
been shown that the resummation of soft-gluon effects can
significantly reduce the scale uncertainties for $gg \rightarrow
ZH$.}. In this paper we take a step towards this goal, as we provide a
fast and flexible way to compute $gg \rightarrow ZH$
at NLO in QCD including the effects of a finite top-quark mass .

We obtain a very reliable approximation of the two-loop virtual corrections by
merging analytic results that are accurate in two complementary
phase-space regions, namely the results of the transverse-momentum expansion of
Ref.~\cite{Alasfar:2021ppe} and the high-energy (HE) expansion of
Ref.~\cite{Davies:2020drs}. In Ref.~\cite{Bellafronte:2022jmo} it has
been shown that this approach can provide a fast evaluation of the
virtual corrections with an accuracy of 1\% or below. In this paper, we
present a complete assessment of the NLO QCD corrections to $gg \rightarrow ZH$ by including the real-emission contributions, which
are related to one-loop diagrams with an additional parton in the
final state.  We use our results to quantify the effects of the
gluon-initiated channel on the hadronic cross section, both at the
inclusive and differential level, and we compare our findings with
previous independent calculations.  Additionally, we discuss the
theoretical uncertainty related to the choice of renormalization
scheme for the top-quark mass. We finally point out a so-far unnoticed
feature of the NLO QCD corrections, as we observe that the
contribution from the class of $2\to 3$ diagrams in which the $Z$ boson
is emitted from an open quark line becomes numerically relevant for
collisions at very high energies, namely for invariant masses in the range $M_{ZH} > 1$ TeV. To the best of our knowledge, the
impact of these diagrams on $ZH$ production over this energy regime has not been discussed in the literature.

This paper is structured as follows: in the next section we set our
notation and we describe the ingredients for the calculation of the
NLO corrections to $gg \rightarrow ZH$; in section \ref{sec:res} we
present our results for the inclusive and differential cross section
and we assess the top-mass scheme uncertainty. In section
\ref{sec:concl} we state our conclusions.

%% file: DGVZ2.tex
\section{The $gg \to ZH$ channel at NLO}

The  cross section for the subprocess $g g \to Z H+X$ in the hadronic
reaction $pp \to Z H+X$ at center-of-mass energy $\sqrt{s}$, can be 
written as
\be
M_{ZH}^2 \frac{d\,\sigma}{d\, M_{ZH}^2}  =  
          \sum_{a,b}\int_0^1 dx_1 dx_2 \,\,f_{a}(x_1,\mu_F^2)\,
         f_{b}(x_2,\mu_F^2) 
\int_0^1 dz~ \delta \left(z-\frac{\tau}{x_1 x_2} \right)
M_{ZH}^2 \frac{d\,\hat\sigma_{ab}}{d\, M_{ZH}^2} \, ,
\label{sigmafull}
\ee
where $M_{ZH}^2$ is the invariant mass of the $Z$-Higgs system,
$\tau = M_{ZH}^2/s$, $\mu_F$ is the factorization scale,
$f_{a}(x,\mu_F^2)$, the parton density of the colliding proton
for the parton of type $a, \,(a = g,q,\bar{q})$ and $\hat\sigma_{ab}$ is the 
partonic cross section for the subprocess $ ab \to ZH +X$ at the partonic
center-of-mass  energy  $\hat{s}=x_1 x_2 s$. The partonic cross section can be 
written in terms of the Born (i.e.~LO) partonic cross section
$\hat{\sigma}^{(0)}$  as:
\be
M_{ZH}^2 \frac{d\,\hat\sigma_{ab}}{d\, M_{ZH}^2}=
\hat\sigma^{(0)}(z \hat{s})\,z \, G_{ab}(z) \, ,
\label{Geq}
\ee
where, up to NLO terms in QCD,
\be
G_{a b}(z)  =  G_{a b}^{(0)}(z) 
          + \frac{\alpha_s (\mu_R)}{\pi} \, G_{a b}^{(1)}(z) \, 
\label{Gas}
\ee
with $\as(\mu_R)$ the strong coupling
constant defined at the renormalization scale $\mu_R$.

The LO  contribution $G_{a b}^{(0)}(z)$ is given by the  $gg \to ZH$ channel only,
 i.e.
\be
G_{a b}^{(0)}(z)  =  \delta(1-z) \,\delta_{ag}\, \delta_{bg} \, .
\ee
The amplitude for $g^\mu_a(p_1)g^\nu_b(p_2)\to Z^\rho(p_3) H(p_4)$ can be written
as
\bea
&&\amp=i \sqrt{2}\frac{\mz G_\mu \as(\mu_R)}{\pi}\delta_{ab}\epsilon^a_\mu(p_1)
\epsilon^b_\nu(p_2)\epsilon_\rho(p_3)\hat{\amp}^{\mu\nu\rho}(p_1,p_2,p_3 ),\\
&&\hat{\amp}^{\mu\nu\rho}(p_1,p_2,p_3 )=\sum_{i=1}^{6}
\mathcal{P}_i^{\mu\nu\rho}(p_1,p_2,p_3 )
\amp_i(\hat{s},\hat{t},\hat{u},\mt,\mh,\mz),
\label{eq:amp}
\eea
where $G_\mu$ is the Fermi constant and
$\epsilon^a_\mu(p_1)\epsilon^b_\nu(p_2)\epsilon_\rho(p_3)$ are the
polarization vectors of the gluons and the $Z$ boson, respectively. The
tensors $\mathcal{P}_i^{\mu\nu\rho}$ are a set of orthogonal
projectors whose expressions can be found in Ref.~\cite{Alasfar:2021ppe}.
The corresponding form factors
$\amp_i(\hat{s},\hat{t},\hat{u},\mt,\mh,\mz)$ are functions of the
masses of the top quark\footnote{We neglect the masses of all quarks except for the top.} ($\mt$), Higgs ($\mh$) and $Z$ ($\mz$) bosons, and of
the partonic Mandelstam variables
\be
\hat{s}=(p_1+p_2)^2,~~ \hat{t}=(p_1+p_3)^2,~~ \hat{u}=(p_2+p_3)^2,
\ee
where $\hat{s}+\hat{t}+\hat{u}=\mz^2+\mh^2$ and we took all the momenta to
be incoming.

The $\amp_i$ form factors can be expanded up to NLO terms as
\be
\amp_{i} = \amp_i^{(0)} + \frac{\as}{\pi} \amp_i^{(1)}
\label{eq:ampexp}
\ee
with
\bea
  \mathcal{A}_{i}^{(0)} &=& \mathcal{A}_{i}^{(0, \triangle) } +
  \mathcal{A}_{i}^{(0, \square) } ,
  \label{eq:ampformu}                 \\
  \mathcal{A}_{i}^{(1)}& = &\mathcal{A}_{i}^{(1, \triangle) } +
  \mathcal{A}_{i}^{(1, \square) } + \mathcal{A}_{i}^{(1, \bowtie) }~,
  \label{eq:ampformd}
\eea
where the LO amplitude can be written in terms of two contributions,
namely the one-loop  triangle diagrams ($\mathcal{A}_{i}^{(0, \triangle) }$) and the one-loop  box diagrams ($\mathcal{A}_{i}^{(0, \square) }$), while
at NLO besides the genuinely two-loop triangles
($\mathcal{A}_{i}^{(1, \triangle) }$) and the two-loop
boxes ($\mathcal{A}_{i}^{(1, \square) }$) also the contribution from reducible
double-triangle diagrams ($\mathcal{A}_{i}^{(1, \bowtie) }$) is present. In the latter we include also diagrams featuring loops of bottom quarks connecting to the $Z$ boson, and we set the bottom mass to zero. 
In Eq.~\eqref{eq:ampformd} the  $\mathcal{A}_{i}^{(1, \triangle) }$ and
$ \mathcal{A}_{i}^{(1, \square)}$ contributions are understood as 
regularized  with respect to the ultraviolet (UV) and infrared (IR)
singularities via the introduction of a counterterm as in
Ref.~\cite{Grober:2017uho}.

The  Born partonic cross section can be written as
\be
\hat{\sigma}^{(0)}(\hat{s})=
\frac{\mz^2 G_\mu^2 \as(\mu_R)^2}{32\, \hat{s}\, \pi^2}
\int {\rm d} \Phi\, \sum_i \left|\amp_i^{(0)}\right|^2,
\ee
where
$\rm{d} \Phi$ is the  two-particle Lorentz-invariant phase space.

The NLO terms $G_{a b}^{(1)}$ in Eq.~(\ref{Gas}) include, besides the $gg$ channel, also the $2 \to 3$
processes $gq \rightarrow Z Hq$, $\bar{q}g \rightarrow Z H \bar{q}$
and  $q \bar{q} \rightarrow Z H g$. 
The NLO contribution to the $gg$ channel involves the two-loop
virtual corrections to $g g \rightarrow Z H$ discussed above and one-loop real
corrections from $ gg \to Z H  g$. As well known, the individual
contributions are IR divergent while their sum is finite. In
this work, at the level of cross section, the IR singularities in all channels
were treated via the dipole subtraction method \cite{Catani:1996vz}.
The outcome of this procedure for the $gg$ channel can be summarized as follows
\bea
G_{g g}^{(1)}(z) & = & \delta(1-z) \left[C_A \, \frac{~\pi^2}3 
 \,+ \beta_0 \, \ln \left( \frac{\mu_R^2}{\mu_F^2} \right) \,+ 
\Cnlo  \right]  \nn \\[1mm]
&+ & 
  P_{gg} (z)\,\ln \left( \frac{\hat{s}}{\mu_F^2}\right) +
    C_A\, \frac4z \,(1-z+z^2)^2 \,{\cal D}_1(z) +  {\cal R}_{gg}  \, , 
\label{eq:real}
\eea
where  $C_A =N_c$  ($N_c$ being
the number of colors), $\beta_0 = 11/6\, C_A - 1/2\, N_f\,C_F \,T_R $ ($N_f$
being the number of active flavors, $ C_F = (N_c^2-1)/(2 N_c)$ and $T_R=1/2$)
is the one-loop $\beta$-function
of the strong coupling in the SM,  $P_{gg}$ is the LO Altarelli-Parisi
splitting function
\be
P_{gg} (z) ~=~2\,  C_A\,\left[ {\cal D}_0(z) +\frac1z -2 + z(1-z) \right]
\label{Pgg} \, ,
\ee 
and
\be
{\cal D}_i (z) =  \left[ \frac{\ln^i (1-z)}{1-z} \right]_+  \label {Dfun} \, 
\ee
where the plus distribution is used.

The first line of Eq.~\eqref{eq:real} displays the two-loop virtual
contribution, with 
\be
\Cnlo = 
\,\frac{\int  {\rm d} \Phi\, 2\,\sum_i \text{Re} \left[
\amp_{i}^{(0)} \, \left(\mathcal{A}_{i}^{(1)}\right)^* \right] }{
  \int {\rm d} \Phi \,
  \sum_i \left|\amp_i^{(0)}\right|^2}  \, .
\label{nloC}
\ee
In the second line of Eq.~\eqref{eq:real}, the term ${\cal R}_{gg}$
contains  the integration over the three-particle phase space of
the $gg \to ZHg$ squared amplitude minus two dipole
subtractions\footnote{All the squared matrix elements are understood as averaged
over the initial-state spin, helicity and color and summed over the final ones.}
which cure the IR singularities when the final-state gluon becomes soft or
becomes collinear with either of the initial-state gluons. According to
our normalization (see Eqs.(\ref{Geq}, \ref{Gas})) ${\cal R}_{gg}$ is
obtained dividing the latter quantity by 
$\as(\mu_R) \sigma^{(0)}(z \hat{s})\,z/\pi$. 

The other NLO contributions to $G_{a b}$, i.e. the $gq \rightarrow Z H q,\,
\bar{q}g \to Z H \bar{q}$ and $q \bar{q} \to Z H g$ channels, require
a little discussion. The LO contribution in the hadronic reaction
$pp \to ZH+X$ is given by the tree-level Drell-Yan-like process
$q \bar{q} \to Z^* \to Z H$. The NLO real contribution, ${\cal O}(\as)$,
to this channel includes the tree-level process
$q \bar{q} \to Z H g$ and the crossed channels. The interference between
 these tree-level diagrams and their one-loop corrections
 is usually considered as an  NNLO correction, ${\cal O}(\as^2)$, to the
 Drell-Yan-like contribution and taken into account in the NNLO evaluation of
 $pp \to ZH+X$.\\ 
 The  NLO real corrections to the $ gg \to ZH$ process
  are formally an $\text{N}^3$LO contribution,  ${\cal O}(\as^3)$,
  to the $pp \to ZH+X$ reaction. In this paper we identify  them as the square
  of the one-loop diagrams containing a closed fermion loop to which a Higgs
  or a $Z$ boson, or both particles, are attached in the
 $gq \rightarrow Z H q,\, \bar{q}g \to Z H \bar{q}$ and $q \bar{q} \to Z H g$
  channels. It should be remarked that our definition of the NLO contribution
  to $ gg \to ZH$ process differs from the one employed in
  Refs.~\cite{Wang:2021rxu,Chen:2022rua}. In those references, diagrams in which a $Z$ boson is emitted from an open fermion line, which we denote as $Z$-\emph{radiated diagrams}  for brevity (see subsection \ref{real}), were not taken into account
  because they were assigned to uncalculated $\text{N}^3$LO corrections to
  the Drell-Yan-like contribution.

The contribution of the $q g \to ZH q$ channel, and 
similarly for $\bar{q} g \to ZH \bar{q}$,  can be
written as:
\be
G_{q g}^{(1)}(z) =  P_{gq}(z) \left[ \ln(1-z) + 
  \frac12 \ln \left( \frac{\hat{s}}{\mu_F^2}\right) \right] +
\frac12 C_F z + {\cal R}_{qg} \,,
\label{qgzhq}
\ee
where
\be
P_{gq} (z) ~=~  C_F \,\frac{1 + (1-z)^2}z~,
\ee
and ${\cal R}_{qg}$ is obtained from the integration over the three-particle
phase space of the squared
amplitude of the process minus one dipole subtraction, which cures the IR
singularity
when the final quark becomes collinear to the initial quark,
normalized to $\as(\mu_R) \sigma^{(0)}(z \hat{s})\,z/\pi$.

Finally, the $q \bar{q}  \to ZH g$ channel is IR safe.  Its contribution can be written as
\be
G_{q \bar{q}}^{(1)}(z) =   {\cal R}_{q \bar{q}} \, ,
\label{qqzhg}
\ee
where ${\cal R}_{q \bar{q}}$ is given by the square of the one-loop diagrams
integrated over the phase space and  normalized to
$\as(\mu_R) \sigma^{(0)}(z \hat{s})\,z/\pi$.

In the following subsections we discuss the method used for the
calculation and the implementation of the various contributions to
Eq.~(\ref{Gas}). We evaluated our results using different renormalization schemes for the top-quark mass, namely the on-shell (OS) and the modified minimal subtraction ($\MSbar$) scheme.  In particular, for evaluating the top  mass in the $\MSbar$ scheme, we first convert the OS mass to $m_t^{\MSbar}(\mu_t=m_t^{\textrm{OS}})$ using the three-loop relation\cite{Melnikov:2000qh},
and then run it at three-loop order \cite{Chetyrkin:1997dh} numerically to the indicated scale $\mu_t$.

\input{realcorrection}

%% file: realcorrection.tex
\subsection{Virtual Corrections}
\label{virtual}
Concerning the two-loop form factors $\mathcal{A}_i^{(1)}$ in
Eq.~(\ref{nloC}), the contributions from $\mathcal{A}_i^{(1,
  \triangle)}$ and $\mathcal{A}_i^{(1, \bowtie)}$, as well as the LO contribution, were evaluated
in exact top-mass dependence using the results available from
Ref.~\cite{Alasfar:2021ppe}. Instead, the two-loop box integrals
associated to $\mathcal{A}_i^{(1, \square)}$ were evaluated combining
two analytic approximations corresponding to different kinematical
regimes, and here we briefly recall the main features of this
approach, described in detail in Ref.~\cite{Bellafronte:2022jmo}.

The box integrals depend on five scales, namely $m_Z$, $m_H$, $m_t$
and the kinematic variables $\hat{s}$ and $\hat{t}$, where the latter
can be traded for the transverse momentum of the final-state
particles, $p_T$. In the $p_T$ expansion of
Refs.~\cite{Bonciani:2018omm, Alasfar:2021ppe} it is assumed that the scales associated
to $m_Z, m_H$ and to $p_T$ are small compared to the scales set by
$\hat{s}$ and $m_t$. Under this assumption, the box integrals are
expanded in ratios of small over large scales, and the resulting
simplified integrals are written as linear combinations of 52 master
integrals (MI) using Integration-by-Parts (IBP) identities obtained
with $\texttt{LiteRed}$ \cite{Lee:2013mka, Lee:2012cn}. On the other hand, in the
high-energy expansion used in Ref.~\cite{Davies:2020drs} the two-loop
box integrals are first expanded in terms of small $m_Z$ and $m_H$,
then an IBP reduction is performed on the expanded integrals; the
resulting MIs are further expanded in the limit $m_t^2 \ll \hat{s},
|\hat{t}|$ and expressed in terms of harmonic polylogarithms.

In Ref.~\cite{Bellafronte:2022jmo} it has been shown that, in the
forward kinematic regime\footnote{Here and in
Ref.~\cite{Bellafronte:2022jmo} it is assumed that the form factors
are (anti)symmetric under $\hat{t} \leftrightarrow \hat{u}$. As a
consequence, the discussion above can be adapted to the backward
regime simply replacing $|\hat{t}|$ with $|\hat{u}|$.} defined by
$|\hat{t}| \leq |\hat{u}|$, the results of the $p_T$ expansion are
accurate in the phase-space region $|\hat{t}|\lesssim 4 m_t^2$, while
the HE expansion is accurate in the complementary region
$|\hat{t}|\gtrsim 4 m_t^2$. If the convergence of both expansions is
improved using Pad\'e approximants, then the combination of the
results allows to approximate the exact cross section with an accuracy
of 1\% or below over the whole phase space. This procedure has been
implemented in a {\tt FORTRAN} program, which uses the  Pad\'e-improved HE expansion for phase-space points such that $|\hat{t}| > 4m_t^2$ and $|\hat{u}|> 4m_t^2$, and the Pad\'e-improved  $p_T$
expansion for all the remaining phase-space points. 
When we compute the results using the $\MSbar$ scheme for the top-mass renormalization, the form factors $\mathcal{A}_i^{(1,
  \triangle)}$ and $\mathcal{A}_i^{(1,
  \square)}$ need to be shifted by a quantity that is defined in Eqs.(14,15) in Ref.~\cite{Bellafronte:2022jmo}. This shift is applied to the form factors before the construction of the relative Pad\'e approximants.

We use $\texttt{handyG}$ \cite{Naterop:2019xaf} for the
evaluation of polylogarithms and the routine of
Ref.~\cite{Bonciani:2018uvv} for the evaluation of the two elliptic
integrals occurring in the $p_T$ expansion results. As a result, the
average timing for evaluating one phase-space point using the Pad\'e-improved
HE expansion is around 0.004 s, while using the Pad\'e-improved $p_T$ expansion the timing
ranges from 0.02\,s to 0.09\,s. For comparison, the evaluation of a phase-space point for the virtual terms using the results of a small-mass expansion requires on average 2 s \cite{Wang:2021rxu}. 

\subsection{Real Corrections}
\label{real}
\begin{figure}
    \begin{subfigure}{0.24\textwidth}
        \includegraphics[width=\textwidth]{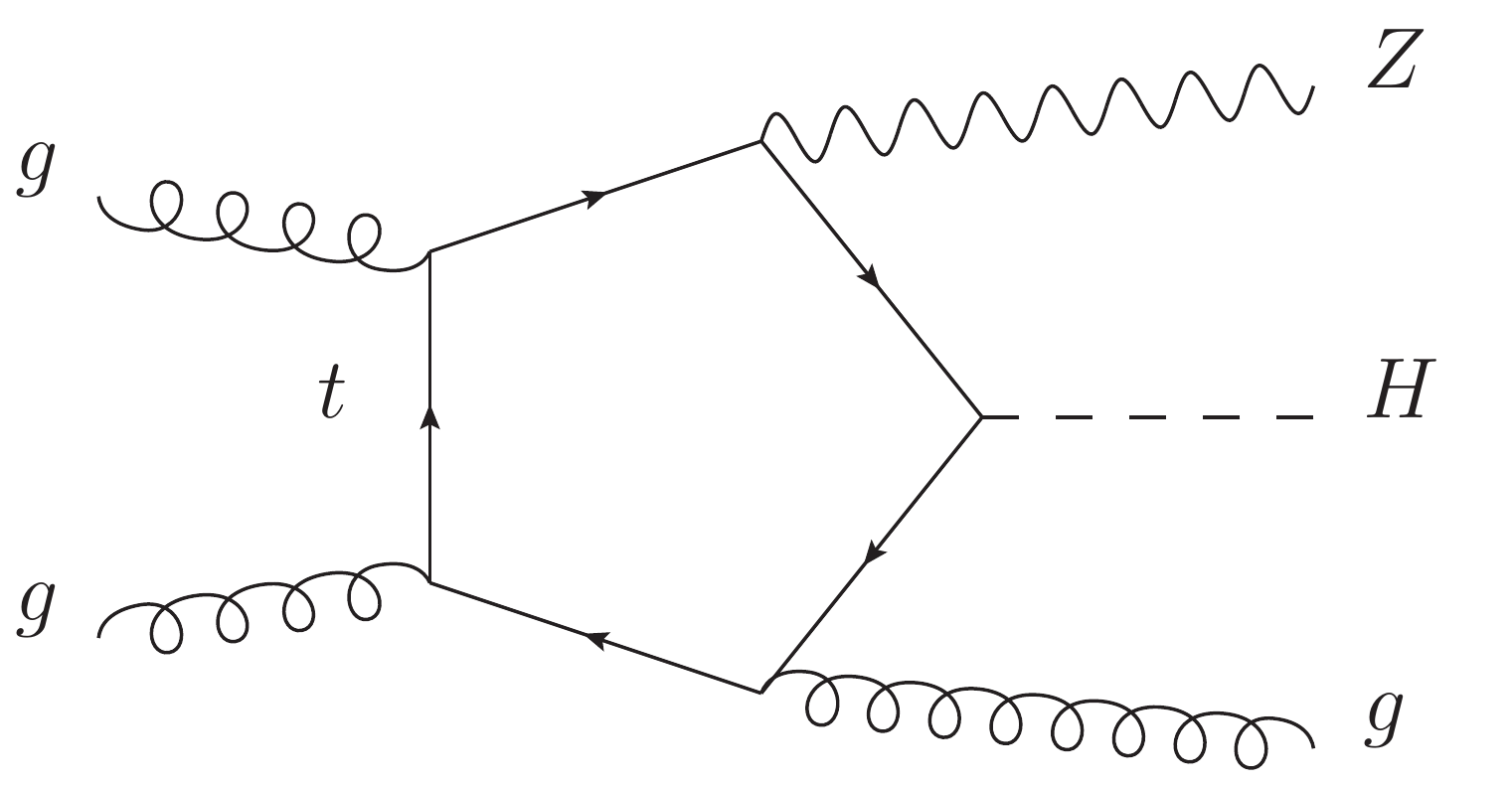}
        \caption{}
    \end{subfigure}
    \begin{subfigure}{0.24\textwidth}
        \includegraphics[width=\textwidth]{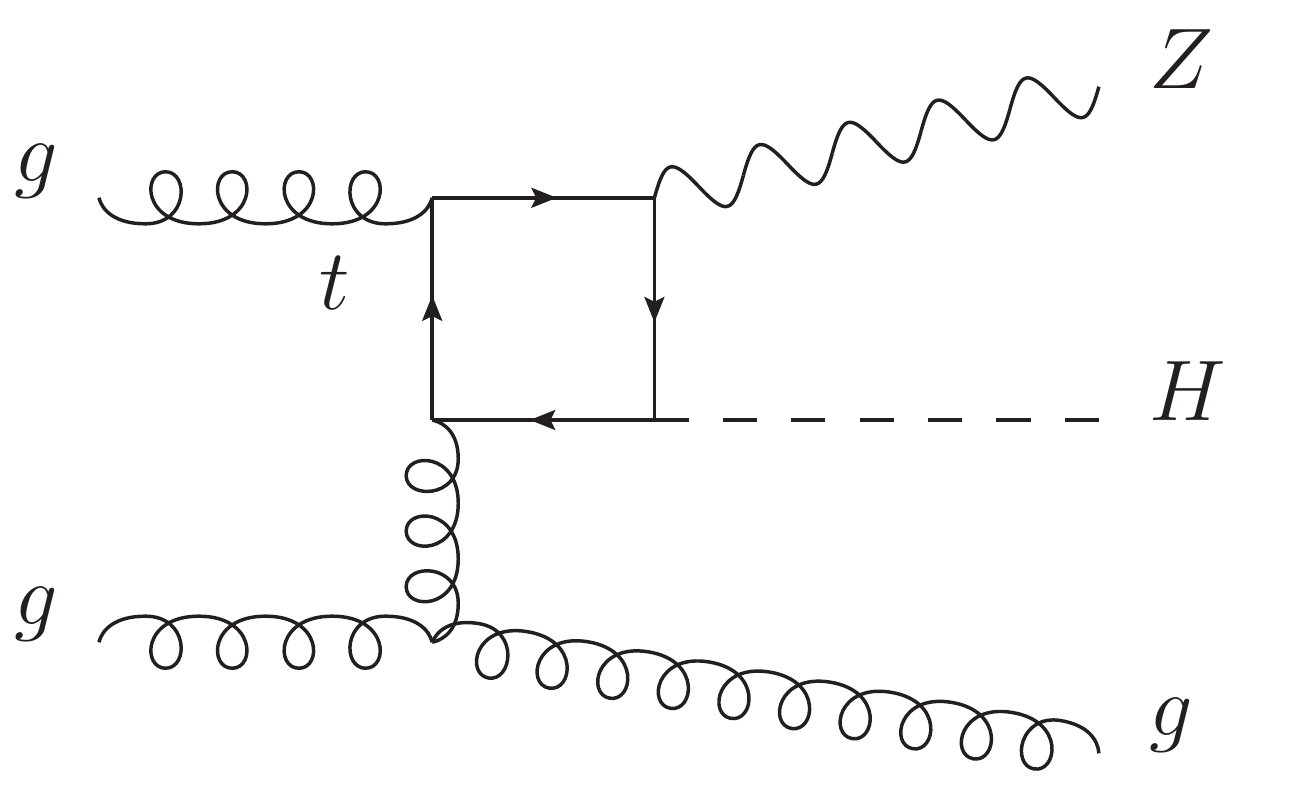}
        \caption{}
    \end{subfigure}
    \begin{subfigure}{0.24\textwidth}
        \includegraphics[width=\textwidth]{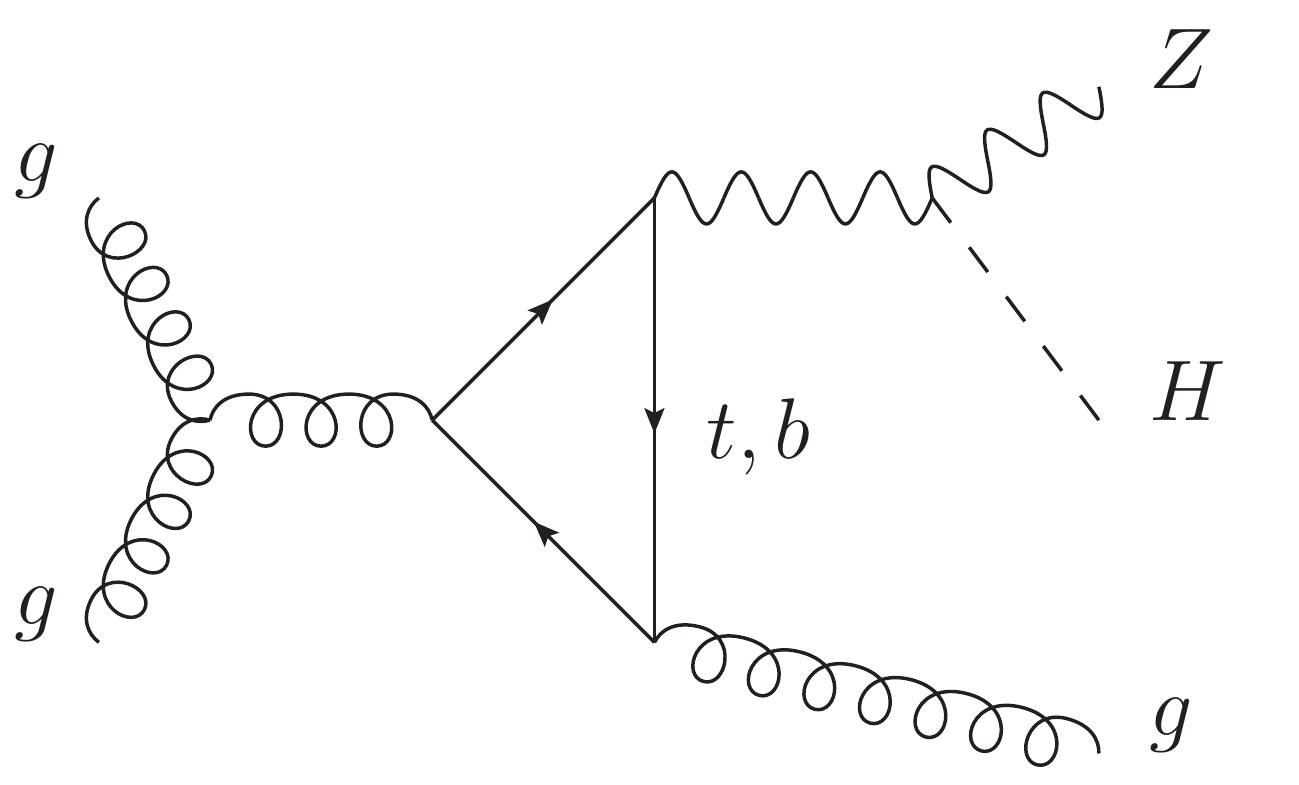}
        \caption{}
        \label{fig:ggtrizh}
    \end{subfigure}
    \begin{subfigure}{0.24\textwidth}
        \includegraphics[width=\textwidth]{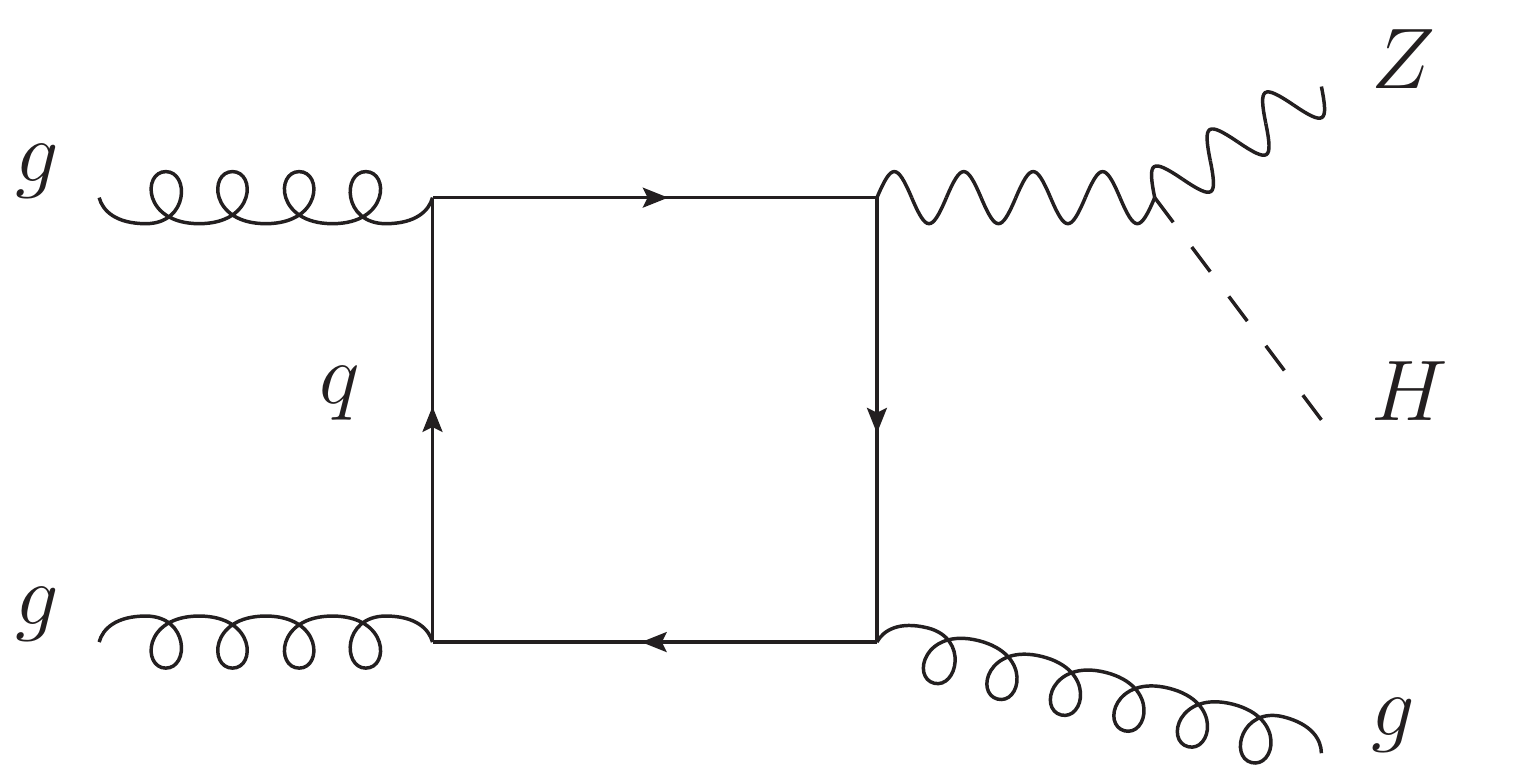}
        \caption{}
    \end{subfigure}
    \caption{Representative Feynman diagrams for the $gg\to ZHg$ process.} 
    \label{fig:ggzhr}
\end{figure}

For the $gg\to ZHg$ process, we adopt {\tt Recola2} \cite{Actis:2012qn,Denner:2017wsf} to compute the one-loop matrix element,
and we cross checked the result with {\tt MadGraph5\_aMC@NLO} \cite{Alwall:2014hca}.
We include all diagrams with massive and massless closed quark loops. Some  representative diagrams are shown in Fig.~\ref{fig:ggzhr}.
We note that in {\tt Recola2}, the value of the top-quark mass cannot be changed after process initialization,
and hence the process needs to be reinitialized each time if a dynamical top mass is adopted (which is the case if a dynamical scale is chosen in the $\MSbar$ top-mass renormalization scheme, see the next section).
As a result, the average time to compute one phase-space point increases from 0.2 s to 1.0 s.

\begin{figure}[ht]
    \centering
    \begin{subfigure}{0.24\textwidth}
        \includegraphics[width=\textwidth]{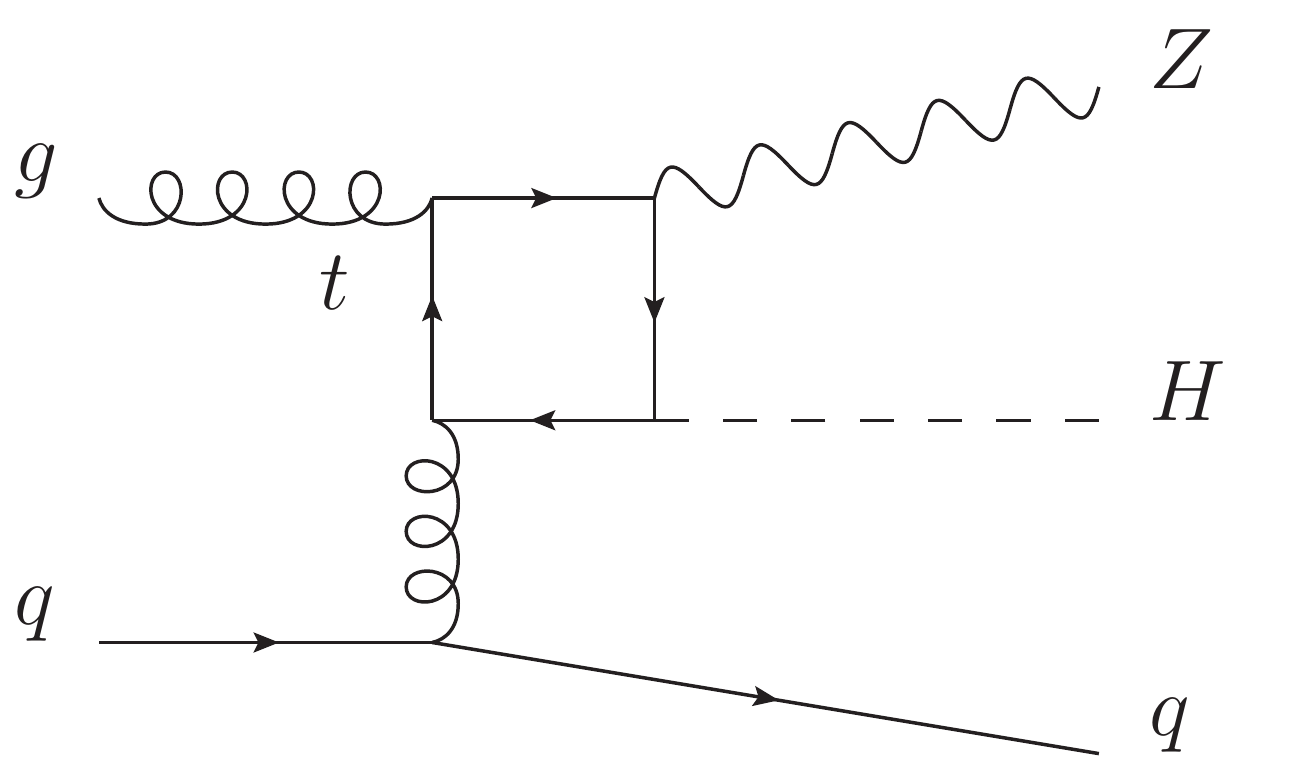}
        \caption{}
        \label{fig:real-qg-hz}
    \end{subfigure}
    \begin{subfigure}{0.24\textwidth}
        \includegraphics[width=\textwidth]{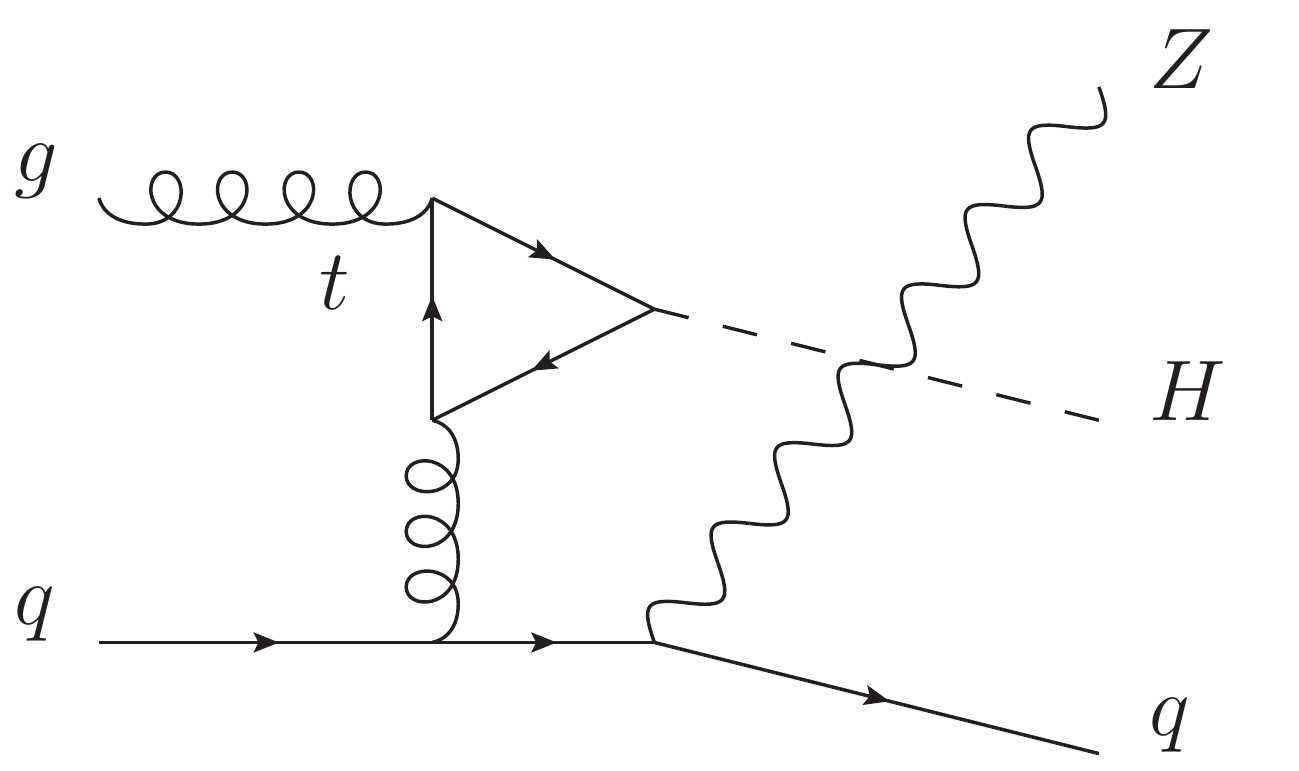}
        \caption{}
        \label{fig:real-qg-zrad}
    \end{subfigure}
    \begin{subfigure}{0.24\textwidth}
        \includegraphics[width=\textwidth]{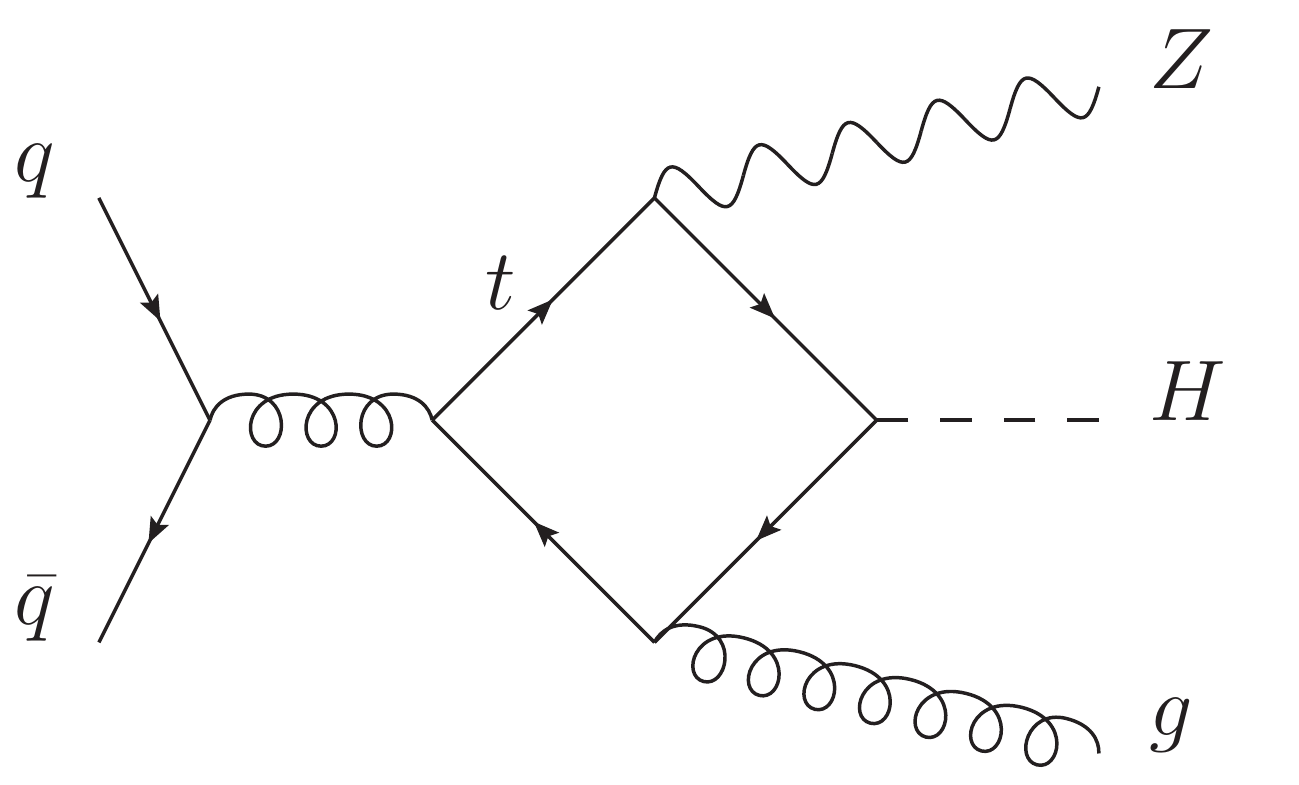}
        \caption{}
        \label{fig:real-qq-hz}
    \end{subfigure}
    \begin{subfigure}{0.24\textwidth}
        \includegraphics[width=\textwidth]{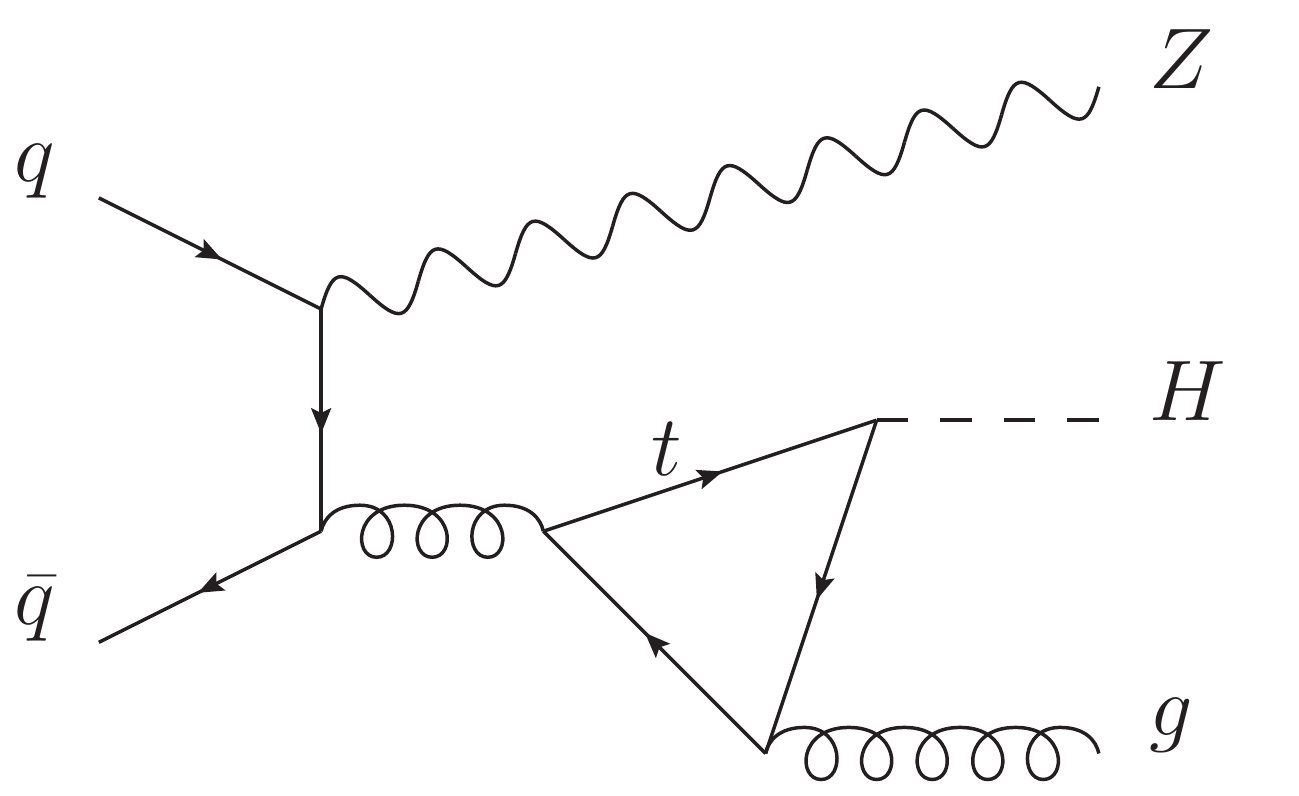}
        \caption{}
        \label{fig:real-qq-zrad}
    \end{subfigure}
    \caption{Representative Feynman diagrams for the $qg$ and $q\bar{q}$ channels. In (b) and (d) examples of $Z$-radiated diagrams (see Sec.~\ref{sec:res}) are depicted.}
    \label{fig:qgqqzhr}
\end{figure}

For $qg\to ZHq$, and $q\bar{q}\to ZHg$, the one-loop matrix elements are computed by {\tt MadGraph5\_aMC@NLO}, 
where we implement a filter to exclude diagrams without a closed fermion loop.
In other words, we include two classes of Feynman diagrams:
in the first class, examples of which are shown in Figs. \ref{fig:real-qg-hz} \ref{fig:real-qq-hz}, both the $Z$ boson and Higgs boson are attached directly or indirectly (i.e.~by connecting to an intermediate virtual boson, similarly to Fig.~\ref{fig:ggtrizh}) to a closed quark loop,
while in  the second class (as shown in Fig.~\ref{fig:real-qg-zrad} \ref{fig:real-qq-zrad}) the Higgs boson is attached to a closed quark loop,
but the $Z$ boson is radiated from an open fermion line. 
We note that both types of diagrams can interfere with tree-level diagrams,
hence produce $\mathcal{O}(\as^2)$ contributions.
Such contributions were studied in detail\footnote{They belong to the classes $R_I$ and $R_{II}$ for the \emph{top-mediated} terms considered in Ref.~\cite{Brein:2011vx}.} in Ref.~\cite{Brein:2011vx}  and they were considered as part of the NNLO corrections to $pp \to ZH$.
On the other hand,
in this paper we compute the square of those diagrams,
corresponding to $\mathcal{O}(\as^3)$ contributions that we consider as NLO corrections to $gg\to ZH$.

%% file: Res.tex
\section{Results}
\label{sec:res}
In this section, we present our numerical results for a center-of-mass energy $\sqrt{s}=13\text{ TeV}$.
We adopt the following input parameters: $m_t^{\textrm{OS}}=172.5\,\mathrm{ GeV},\;m_W=80.385\,\mathrm{GeV},\;m_Z=91.1876\,\mathrm{ GeV},\;m_H=125\,\mathrm{ GeV},G_{\mu}=1.1663787\times10^{-5}\,{\mathrm{ GeV}^{-2}}$.
We adopt the \newline
{\tt NNPDF31\_nnlo\_as\_0118} \cite{NNPDF:2017mvq} parton distribution functions in a five flavour scheme.

\subsection{Inclusive Cross Section}
In Table \ref{tab:totxs}, we show the total cross section at LO and NLO adopting different top-quark-mass renormalization schemes, i.e.~OS and  $\MSbar$ with different scale choices.
We fix the central value of the renormalization and factorization scales to be $\mu_C=M_{ZH}/2$. The scale uncertainty is obtained from the envelope of a 7-point variation of the central scale according to $(\mu_R/\mu_C,\mu_F/\mu_C)=(1,1),(1,\frac{1}{2}),(1,2),(\frac{1}{2},\frac{1}{2}),(\frac{1}{2},1),(2,1),(2,2)$. 

\begin{table}
    \centering
    \begin{tabular}{c|c|c|c|c|c}
        Top-mass scheme & LO [fb] & $\sigma_{LO}/\sigma_{LO}^{OS}$ & NLO [fb] & $\sigma_{NLO}/\sigma_{NLO}^{OS}$ & $K=\sigma_{NLO}/\sigma_{LO}$\\
        \hline
        On-Shell & $64.01_{-20.3\%}^{+27.2\%}$ & - & $118.6_{-14.1\%}^{+16.7\%}$ & - & 1.85\\
        \hline
        $\MSbar,\mu_t=M_{ZH}/4$ & $59.40_{-20.2\%}^{+27.1\%}$ & 0.928 & $113.3_{-14.5\%}^{+17.4\%}$ & 0.955 & 1.91\\
        $\MSbar,\mu_t=m_t^{\MSbar}(m_t^{\MSbar}) $ & $57.95_{-20.1\%}^{+26.9\%}$ & 0.905 & $111.7_{-14.6\%}^{+17.7\%}$ & 0.942 & 1.93 \\
        $\MSbar,\mu_t=M_{ZH}/2$ & $54.22_{-20.0\%}^{+26.8\%}$ & 0.847 & $107.9_{-15.0\%}^{+18.4\%}$ & 0.910 & 1.99\\
        $\MSbar,\mu_t=M_{ZH}$ & $49.23_{-19.9\%}^{+26.6\%}$ & 0.769 & $103.3_{-15.6\%}^{+19.6\%}$ & 0.871 & 2.10\\
    \end{tabular}
    \caption{Total cross section at LO and NLO with full top-quark mass dependence using different top-quark-mass renormalization schemes. The central value of the renormalization and factorization scales is fixed to be $\mu_R=\mu_F=M_{ZH}/2$. Scale uncertainties are taken from a 7-point scale variation.
    }\label{tab:totxs}
\end{table}

We find that the NLO corrections are large for each choice of the top-mass renormalization scheme, 
with an approximate $K$-factor, $K=\sigma_{NLO}/\sigma_{LO}$, of around 2.
Moreover, the relative size of the scale uncertainties is essentially the same regardless of the top-mass renormalization scheme.
We note that going from LO to NLO the relative size of the scale uncertainties is reduced by a factor of about $2/3$.
The OS scheme leads to the  largest value of the total cross section both at LO and NLO,
while in the $\MSbar$ scheme for $\mu_t=M_{ZH}$ the smallest cross section value is obtained.
At LO, the difference between these two schemes amounts to about 23\%,
while it decreases to 13\% at NLO.

We notice that our OS results are about 20\% larger at LO and 14\% larger at NLO than those of Ref.~\cite{Chen:2022rua} (see Table 1 therein). This discrepancy is mainly due to the different choice for $\mu_C$, and it is related only in a minor way to the additional diagrams included in our calculation and to the different input parameters adopted.
To verify this, we have computed our results including the same diagrams and adopting the same input parameters as in Ref.~\cite{Wang:2021rxu} (which is in accordance with Ref.~\cite{Chen:2022rua}) and we have found an agreement at the level of the Monte Carlo error. Furthermore, when we consider the relative importance of the scale uncertainties, we observe very similar results to Ref.~\cite{Chen:2022rua}.

\subsection{Differential Distributions}
\begin{figure}[t]
\centering
\begin{subfigure}{.5\textwidth}
  \centering
  \includegraphics[scale=0.6]{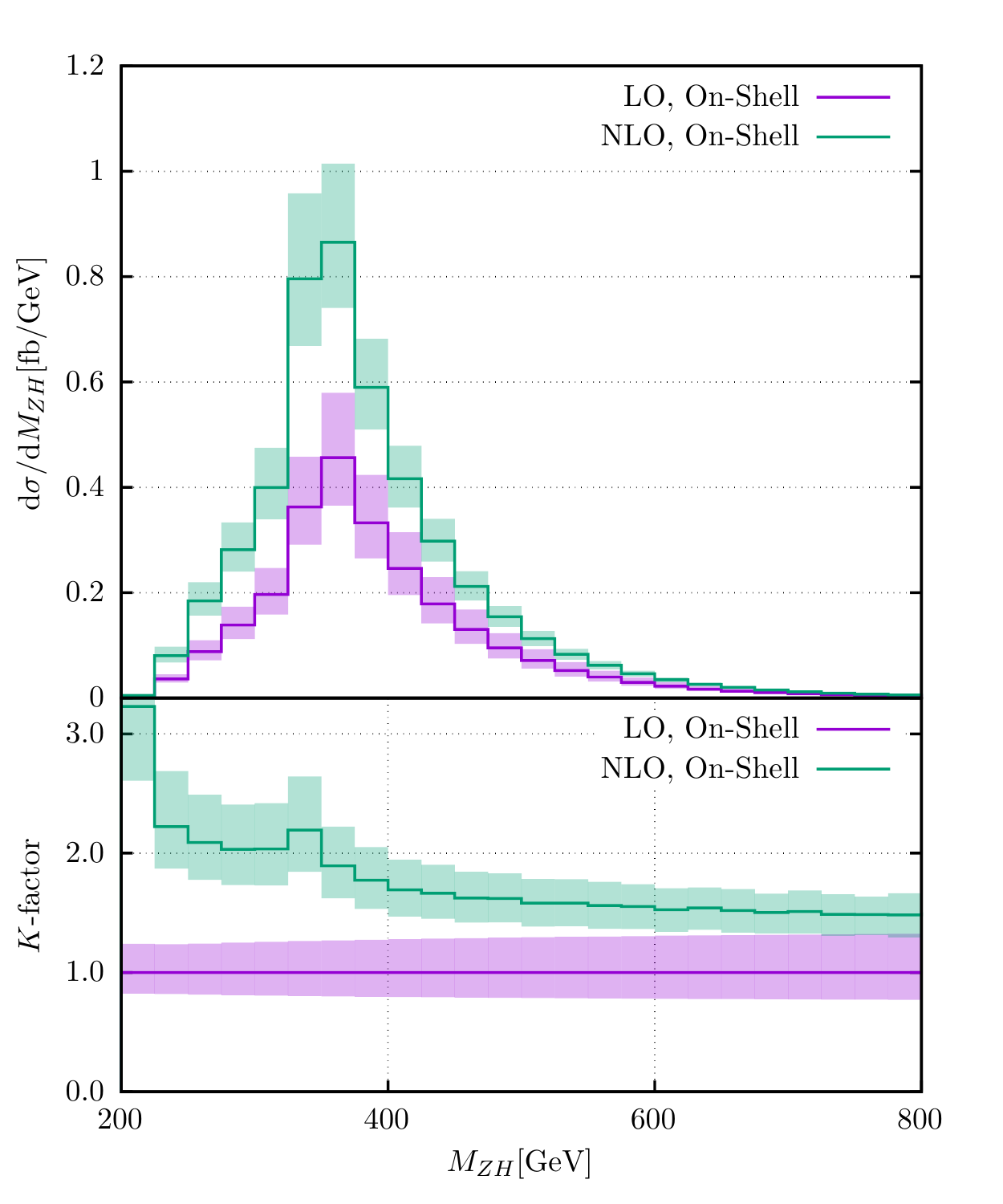} 
  \caption{}
\end{subfigure}%
\begin{subfigure}{.5\textwidth}
  \centering
  \includegraphics[scale=0.6]{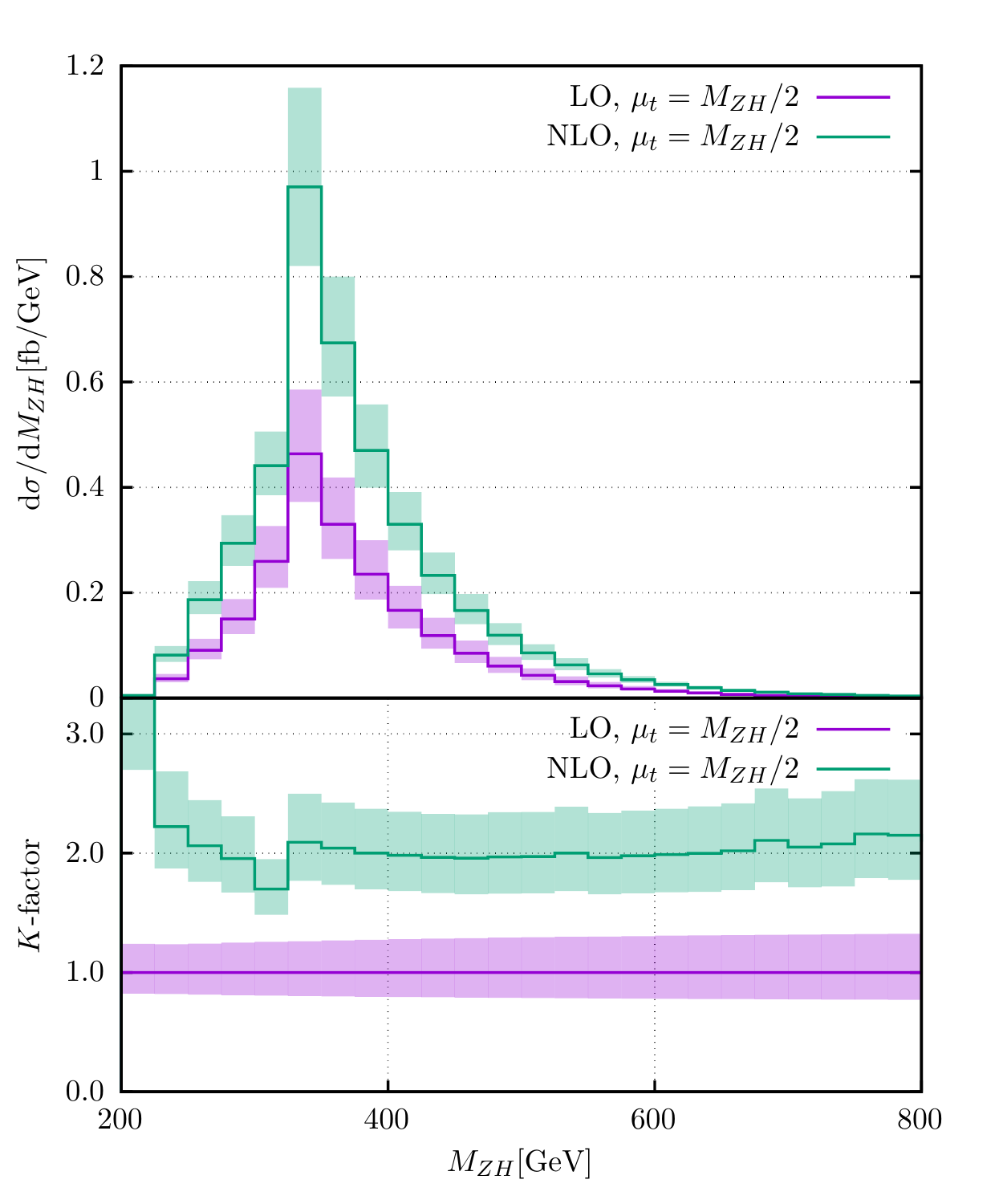} 
  \caption{}
\end{subfigure}\\
\caption{Invariant-mass distribution at LO (magenta) and NLO (green) for the OS scheme (a) and the $\MSbar$ scheme with the choice $\mu_t = M_{ZH}/2$ (b). The scale uncertainties are depicted as shaded bands. The lower panels show the $K$-factor.}
\label{fig:dif}
\end{figure}

\begin{figure}[ht]
\begin{subfigure}{.5\textwidth}
  \centering
  \includegraphics[scale=0.6]{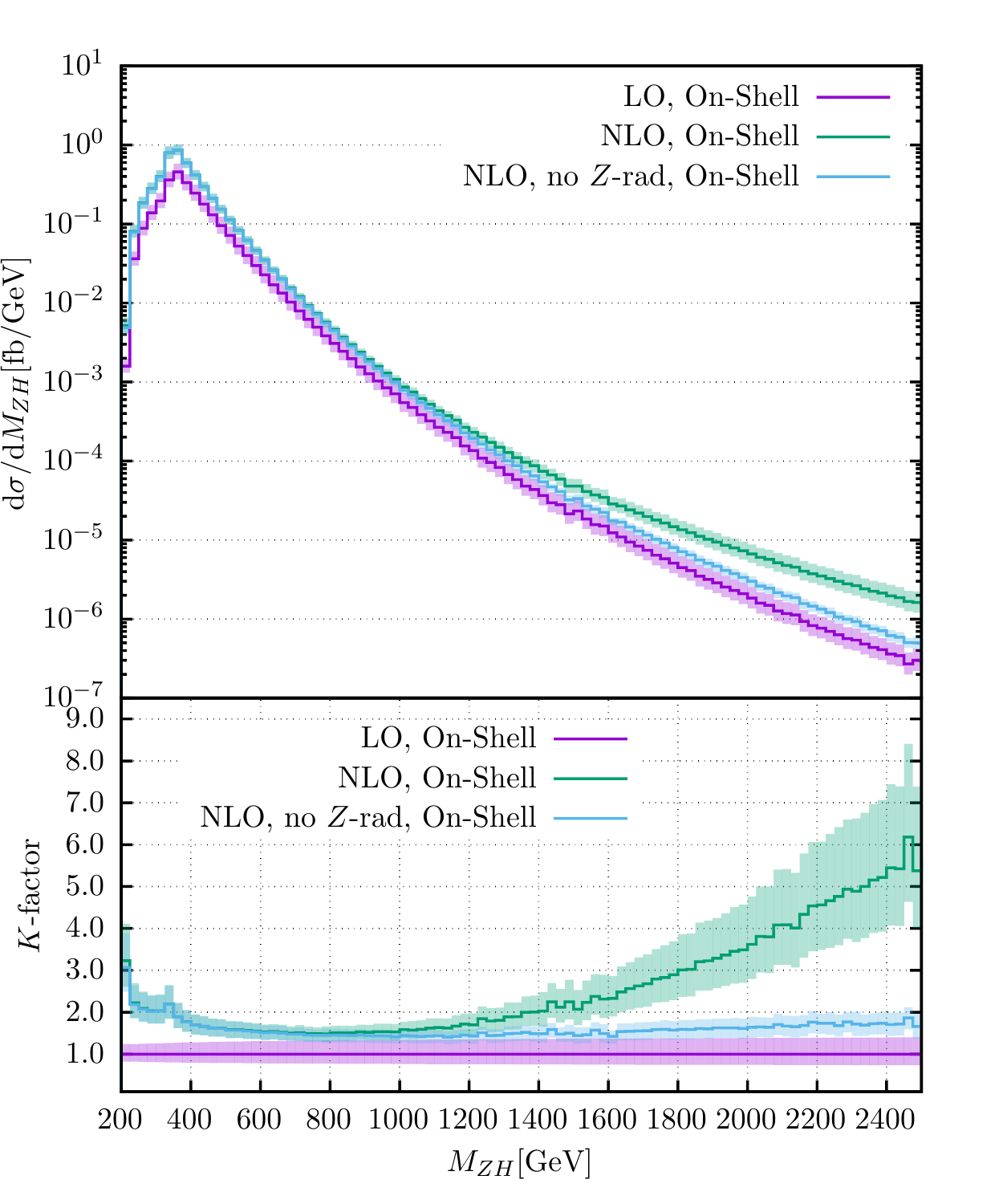} 
  \caption{}
\end{subfigure}%
\begin{subfigure}{.5\textwidth}
  \centering
  \includegraphics[scale=0.6]{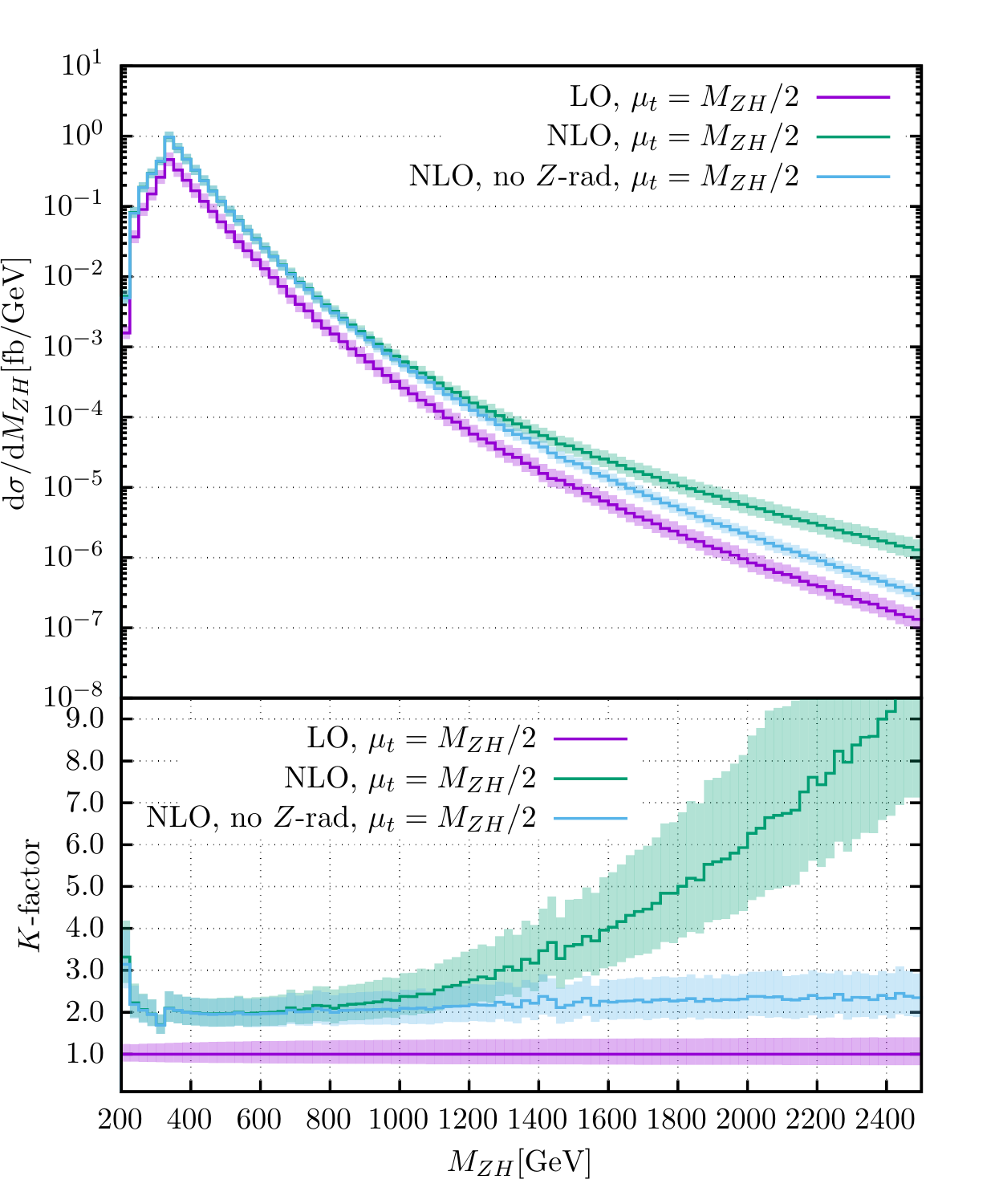} 
  \caption{}
\end{subfigure}
\caption{Invariant-mass distribution at LO (magenta) and NLO (green) for the OS scheme (a) and the $\MSbar$ scheme (b) for a wide $M_{ZH}$ range. The NLO results in which the $Z$-radiated diagrams are excluded are shown in blue. The lower panels show the $K$-factor.}
\label{fig:diflg}
\end{figure}

In Fig.~\ref{fig:dif}, we plot the $M_{ZH}$ distribution in both the OS scheme (\ref{fig:dif}(a)) and the $\MSbar$ scheme with $\mu_t=M_{ZH}/2$ (\ref{fig:dif}(b)) in the region
$M_{ZH}\in [200,800]$ GeV.
In both schemes, the $K$-factor is about 3 in the $ZH$ threshold region,
then it decreases as $M_{ZH}$ increases.
In the top-pair threshold region ($M_{ZH}\sim 2\,m_t$),
the OS scheme gives a peak with $K$-factor slightly above 2,
while the $\MSbar$ scheme shows a small dip followed by a peak instead.
Increasing $M_{ZH}$ to about 800 GeV, the $K$-factor in the OS scheme decreases to about 1.5, while it remains about 2 in $\MSbar$ scheme.

If $M_{ZH}$ is increased to very large values, as shown in Fig. \ref{fig:diflg}, we observe that
the $K$-factor starts to increase rapidly.
At $M_{ZH}=2.5 \,\mathrm{TeV}$, the value of the $K$-factor can reach $\sim6$ in the OS scheme,
and $\sim10$ in the $\MSbar$ scheme.
Such behaviour is due to the inclusion of diagrams where the $Z$ boson is radiated from an open quark line, as in Figs.~\ref{fig:real-qg-zrad} and \ref{fig:real-qq-zrad}. For comparison, in Fig.~\ref{fig:diflg} we
show also the NLO cross section when these contributions are excluded. Indeed, one finds that in the latter case the 
 $K$-factor remains rather flat at high $M_{ZH}$ in both schemes.

\begin{figure}[ht]
\centering
\begin{subfigure}{.49\textwidth}
  \centering
  \includegraphics[width=\textwidth]{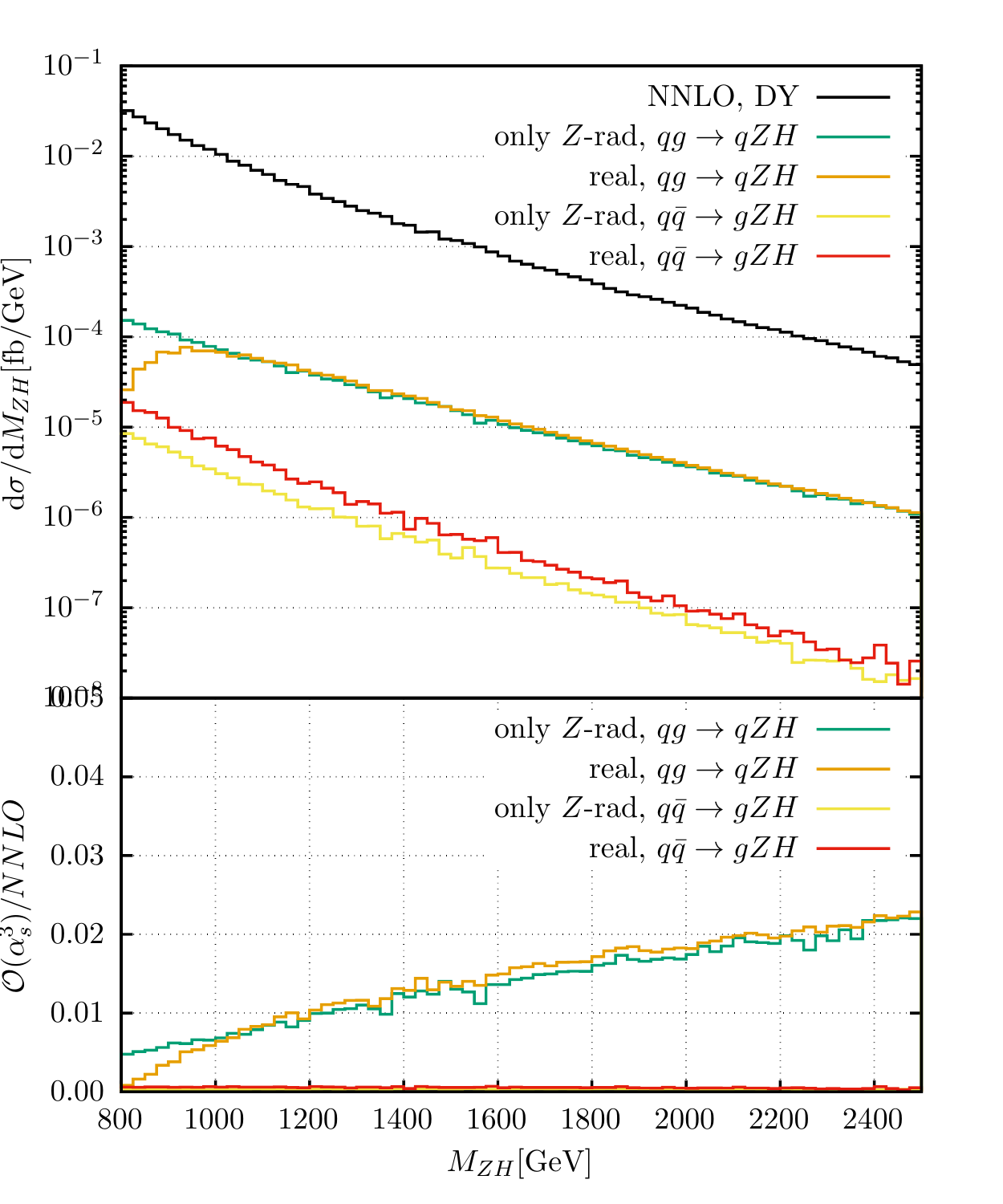} 
  \caption{}
\end{subfigure}
\begin{subfigure}{.49\textwidth}
  \centering
  \includegraphics[width=\textwidth]{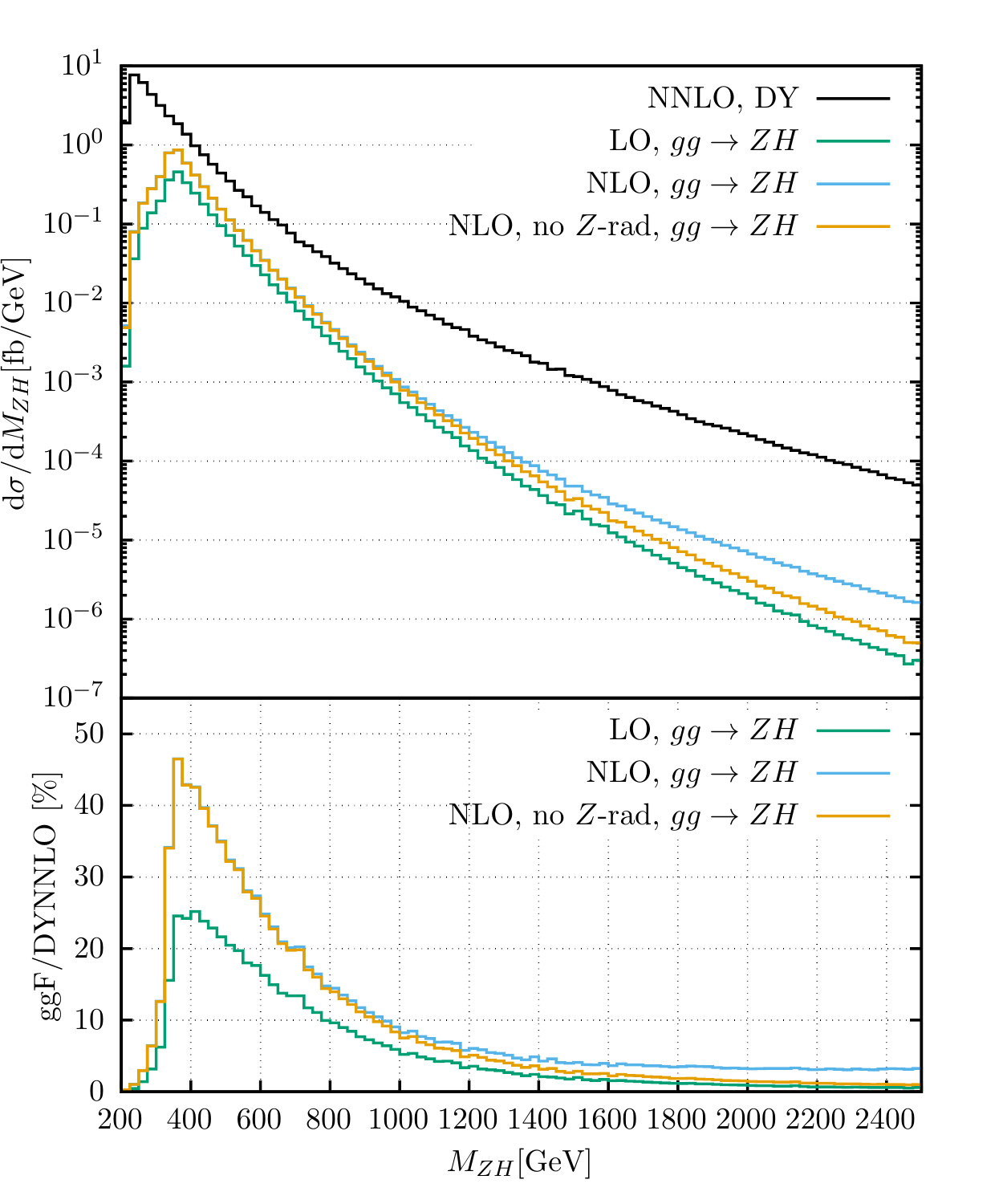} 
  \caption{}
\end{subfigure}
\caption{(a) Various contributions to $ZH$ production at $\mathcal{O}(\as^3)$: the real-emission contributions from the $qg$ and $q\bar{q}$ channel that include all the diagrams considered in this paper are shown as orange and red lines, respectively. The contributions to the same channels including only the $Z$-radiated diagrams are shown as  green ($qg$) and yellow ($q \bar{q}$) lines. The Drell-Yan-like contribution at NNLO is shown as a black line. 
(b) Comparison of the invariant-mass distribution for $gg\to ZH$ at LO and NLO with the Drell-Yan-like contribution at NNLO. For $gg\to ZH$ at NLO, the results with (blue) and without (orange) the contribution from the $Z$-radiated diagrams are shown.} 
\label{fig:Zradeffects}
\end{figure}

To further assess the contribution of these $Z$-radiated diagrams,
in Fig.~\ref{fig:Zradeffects}(a) we show various pieces of the $\mathcal{O}(\as^3)$ corrections to the $pp \rightarrow ZH+X$ cross section: we compare the isolated contribution of the $Z$-radiated diagrams for the $qg$ and $q \bar{q}$ initial states (green and yellow lines, respectively) to the contribution from each of the two partonic channels including all the relevant diagrams (orange and red lines for $qg$ and $q \bar{q}$, respectively).
We can see that in the $qg$ channel when $M_{ZH}>1 \,\mathrm{TeV}$,
the dominant contribution comes from the square of the $Z$-radiated diagrams. We ascribe this feature to the contribution from logarithmic terms of EW origin, of the form $\log^2(m_Z^2/M_{ZH}^2)$, which become large when the typical scale of the process ($M_{ZH}$) and the EW scale (represented by $m_Z$) are very different, see e.g. Ref.~\cite{Ciafaloni:2000df}.
On the other hand, although the $q\bar{q}$ channel includes diagrams where the $Z$ boson is radiated from the
external quark lines,
its size remains negligible with respect to the total cross section, though also there we observe
that the $Z$-radiated diagrams are dominating the respective initial state at high $M_{ZH}$. This suppression can be mainly attributed to the reduced partonic luminosity with respect to the $qg$ channel.
For comparison, in Fig.~\ref{fig:Zradeffects}(a) we also report the size of the Drell-Yan type contribution at NNLO (black line), which we obtained using {\tt vh@nnlo} \cite{Brein:2012ne,Harlander:2018yio} with {\tt MCFM}\cite{Campbell:2011bn,Campbell:2015qma,Boughezal:2016wmq}. 
In the lower panel of Fig.~\ref{fig:Zradeffects}(a) we plot the ratio of the $\mathcal{O}(\as^3)$ corrections computed by us with respect to the NNLO Drell-Yan contribution.
We can see that despite being $\mathcal{O}(\as^3)$, the relative importance of the $Z$-radiated contribution can reach 2\% when $M_{ZH}\sim 2\,\mathrm{TeV}$.

In Fig.~\ref{fig:Zradeffects}(b) we compare our results for $gg\to ZH$ at LO (green line) and NLO (blue line) with the Drell-Yan type contribution (black line).
In the upper panel we show the size of the differential cross section for the various channels, while in the lower panel the ratio of the gluon-fusion with respect to the NNLO Drell-Yan contribution is displayed.
We can see that the gluon-fusion contribution peaks around the top-pair threshold,
which increases its relative size over the Drell-Yan contribution by about 25\% at LO,
and about 45\% at NLO.
The relative size of the gluon-fusion contribution decreases above the top-pair threshold as $M_{ZH}$ increases, and at NLO becomes dominated  by the $Z$-radiated terms for very large values of $M_{ZH}$. In particular, at 2 TeV the latter constitute more than half of the gluon-fusion contribution.
\subsection{Change of Renormalization Scheme}
\begin{figure} [htb]
\centering
\begin{subfigure}{.5\textwidth}
  \centering
  \includegraphics[scale=0.7]{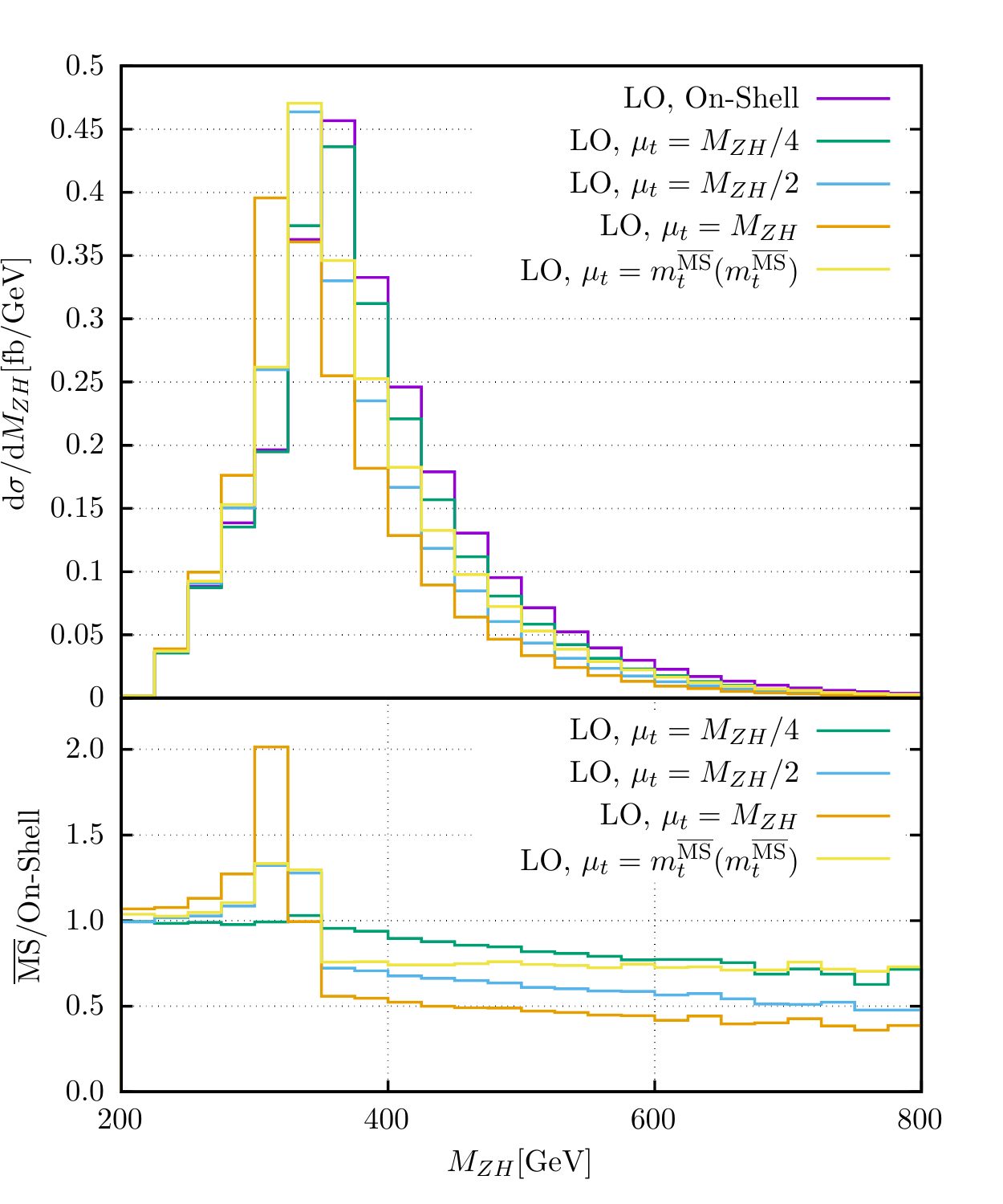} 
  \caption{}
\end{subfigure}%
\begin{subfigure}{.5\textwidth}
  \centering
  \includegraphics[scale=0.7]{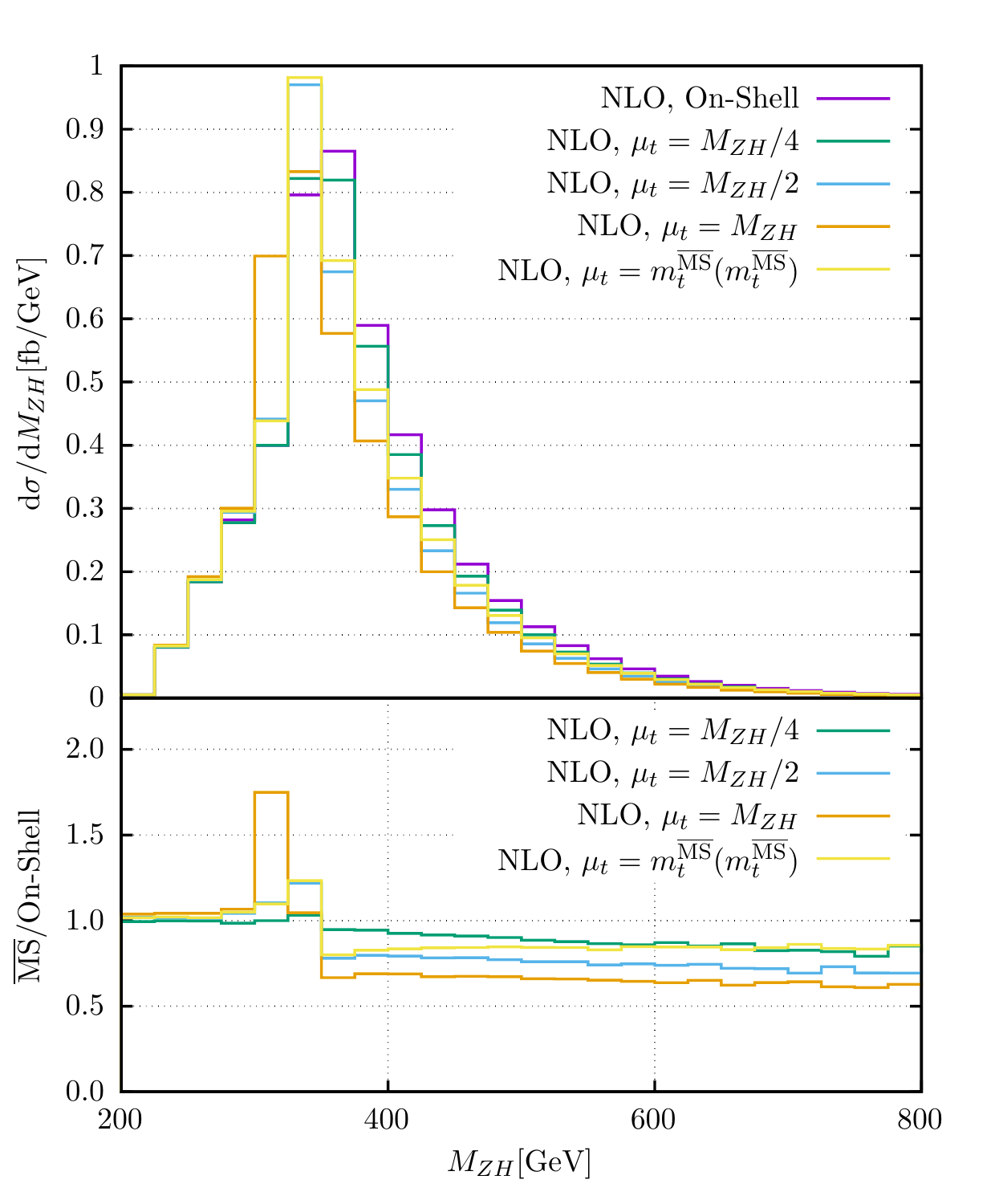} 
  \caption{}
\end{subfigure}
\caption{Invariant-mass distribution at LO (a) and NLO (b) for different choices of the top-mass renormalization scheme. The lower panels show the ratio of the $\MSbar$ results using various scales over the OS results.}
\label{fig:osvsmsbar}
\end{figure}

To assess the impact of the top-quark-mass renormalization scheme at differential level,
we show in Fig.~\ref{fig:osvsmsbar} the differential cross section at LO (\ref{fig:osvsmsbar}(a)) and NLO (\ref{fig:osvsmsbar}(b)) in various top-mass schemes. The lower panel shows the ratio of the $\MSbar$ results with respect to the OS scheme.

Before discussing the results, we note that the value of the top mass can affect the shape of the distributions in two ways: first, in the high-energy regime, it directly controls the overall size of the LO and NLO contributions via the proportionality of the amplitude to the top-Yukawa coupling\footnote{In the low-energy regime $M_{ZH} \lesssim 2 m_t$, where an expansion in the large-$m_t$ limit is accurate, the amplitude goes to a constant, as observed for the first time in Ref.~\cite{Kniehl:1990zu}.}; second, in the low $M_{ZH}$ region the value of $m_t$ shifts the position of the peak associated to the top-pair threshold, $M_{ZH} = 2m_t$. The former effect has been already observed by the authors of Ref.~\cite{Chen:2022rua}, and our results are in agreement when the same scale choices ($\mu_t =M_{ZH}$ and $\mu_t =m_t^{\MSbar}$) and the same invariant-mass range ($M_{ZH} > 400$ GeV) are considered. On the other hand, since we can use our analytical results to investigate the low $M_{ZH}$ region, in Fig.~\ref{fig:osvsmsbar} we are also able to observe the effect of the peak shift on the invariant-mass distribution.
In particular, we can see that the ratio (orange line) between $\mu_t=M_{ZH}$ and the OS scheme is about 2 in the bin $M_{ZH}\in[300,325]\,\mathrm{GeV}$ at LO, and decreases to about 1.75 at NLO. Instead, for $\mu_t=M_{ZH}/4$ (green line) this effect is rather small. Indeed, since renormalization-group evolution predicts that $m_t^{\MSbar}(\mu_t)$ decreases monotonically as $\mu_t$ increases, we expect larger deviations from the OS results when larger values of $\mu_t$ are chosen, see also Ref.~\cite{Baglio:2020ini}.
In the region $M_{ZH}>350\,\mathrm{GeV}$, the OS scheme provides the largest cross section,
while the cross section for the choice $\mu_t=M_{ZH}$ is the smallest.
Across all regions, we can see that going from LO to NLO reduces the difference among different top-mass schemes.

To quantify the size of top-mass renormalization uncertainties,
we follow the procedure adopted in Refs.~\cite{Baglio:2018lrj,Baglio:2020ini}. In this approach, after a binning of the $M_{ZH}$ range is chosen, the OS scheme result is taken as the central value, while for each bin the uncertainty is obtained using the minimal and maximal values from a set of results in different renormalization schemes. 
In particular, for the top mass we considered the set $\left\{m_t^{\text{OS}};~ m_t^{\MSbar} (m_t^{\MSbar});~ m_t^{\MSbar} (M_{ZH}/4) ;~ m_t^{\MSbar} (M_{ZH}/2) ;~ m_t^{\MSbar} (M_{ZH}) \right\}$.
In Table \ref{tab:topunc}, we show our results under different choices of the bin size. Due to the way in which the overall uncertainty is constructed, the latter becomes of course bigger the smaller the bin size is chosen. 
This is though mainly due to the
top-pair threshold region, where the location of the bin can lead to a bigger
or smaller top-mass renormalization uncertainty in dependence on how
well the structure at the top-pair threshold\footnote{For instance, in Ref.~\cite{Bellafronte:2022jmo} a peak dip structure 
for the virtual corrections was observed at the top-pair threshold.} is resolved by the binning, while 
for large invariant masses the OS scheme is always the largest and the scheme using $m_t^{\MSbar}(M_{ZH})$ is the smallest. 
In the very-low invariant-mass region (i.e.~$M_{ZH} \lesssim 275$ GeV) the scheme dependence is small, 
as this region can  be well described by a large mass expansion. We hence expect that if the description of the top-pair threshold region
was improved by a Coulomb resummation, the dependence on the binning of the uncertainty 
would become smaller.

We observe that the uncertainty nearly halves when 
going from LO to NLO for choosing a bin size $\ge 100 \, \mathrm{GeV}$. 
This is similar to what has been observed for $gg\to HH$ in Ref.~\cite{Baglio:2018lrj}. When smaller bin sizes are adopted, the uncertainty does not half but still shows a sizable reduction.

\begin{table}
    \centering
    \begin{tabular}{c|c|c}
        Bin Width [GeV] & LO & NLO \\
        \hline
        1 & $64.01_{-35.9\%}^{+15.6\%}$ & $118.6_{-27.0\%}^{+17.2\%}$\\
        5 & $64.01_{-35.6\%}^{+15.3\%}$ & $118.6_{-24.9\%}^{+14.7\%}$\\
        25 & $64.01_{-33.1\%}^{+14.0\%}$ & $118.6_{-20.8\%}^{+10.9\%}$\\
        100 & $64.01_{-25.3\%}^{+2.0\%}$ & $118.6_{-13.7\%}^{+0.6\%}$\\
        $\infty$ & $64.01_{-23.1\%}^{+0\%}$ & $118.6_{-12.9\%}^{+0\%}$\\
    \end{tabular}
\caption{Inclusive OS results for the LO and NLO $gg \to ZH$ cross sections and relative top-mass-scheme uncertainties. The uncertainties are computed according to Refs.~\cite{Baglio:2018lrj,Baglio:2020ini}.}
\label{tab:topunc}
\end{table}

%% file: Concl.tex
\section{Conclusions}
\label{sec:concl}

In this paper we have presented the evaluation of the QCD corrections at NLO for the $gg$-initiated channel to $ZH$ associated production. The computation of the virtual corrections are based on the approach of Ref.~\cite{Bellafronte:2022jmo}, while for this work we have computed the real corrections and included them in a fast and flexible code. For the inclusive cross section, we have found that the NLO QCD corrections are of the same size as the LO contribution, and they increase the cross section by a factor of about 2. Our findings are in accordance with independent results in the literature \cite{Wang:2021rxu, Chen:2022rua}.
We have also studied the invariant-mass distribution of the $gg\to ZH$ channel at NLO, observing that the perturbative $K$-factor is not flat across the $M_{ZH}$ range, specifically in the region of threshold $ZH$ production and in the very-high-energy tail ($M_{ZH} > 1$ TeV). Furthermore, we have shown that in this latter region the contribution from real-emission diagrams in which the $Z$ boson is emitted from an open quark line is the dominant one, and it causes the $K$-factor to rise up to 10 in the $\MSbar$ scheme. We expect that in $WH$ production an analogous class of $W$-radiated diagrams will give a similar contribution in the high-energy regime. While the focus of our analysis is on LHC phenomenology, this feature could be of interest for studies at future colliders. 

The $gg$-initiated channel is responsible for the larger theoretical uncertainties in the prediction for $ZH$ production, compared to its $WH$ counterpart. We have shown that the inclusion of the NLO corrections bring only a mild reduction of the scale uncertainties, about a factor 2/3, suggesting that more accurate calculations are necessary in order to describe the $gg$-initiated channel at a level that is adequate for  experimental studies in the future. 
Finally, the implementation used for our results allowed us to study for the first time the impact of the uncertainty due to the renormalization scheme for the top quark mass over the whole invariant-mass range: we have found that different choices for the top mass scheme can lead to substantially different results, and we suggest that this uncertainty should be included in refined theoretical predictions.